\documentclass[aps,prb,twocolumn,showpacs,superscriptaddress]{revtex4}
\usepackage{graphicx}
\usepackage{dcolumn}
\usepackage{bm}
\usepackage{color}
\usepackage{amsmath}
\usepackage[export]{adjustbox}

\begin{document}

\newcommand{\bra}[1]{\left\langle#1\right|}
\newcommand{\ket}[1]{\left|#1\right\rangle}
\newcommand{\bracket}[2]{\big\langle#1 \bigm| #2\big\rangle}
\newcommand{\PPV}{\ket{0_{\rm PP}}}
\newcommand{\Tr}{{\rm Tr}}
\renewcommand{\Im}{{\rm Im}}
\renewcommand{\Re}{{\rm Re}}
\newcommand{\MC}[1]{\mathcal{#1}}
\newcommand{\p}{{\prime}}
\newcommand{\pp}{{\prime\prime}}
\newcommand{\ppp}{{\prime\prime\prime}}
\newcommand{\pppp}{{\prime\prime\prime\prime}}
\newcommand{\TK}{T_{\rm K}}

\title{Competition between quantum spin tunneling and Kondo effect}
\author{D. Jacob} 
\email{djacob@mpi-halle.de}
\affiliation{Max-Planck-Institut f\"ur Mikrostrukturphysik, 06120 Halle, Weinberg 2, Germany}
\author{J. Fern\'andez-Rossier}
\affiliation{International Iberian Nanotechnology Laboratory (INL), 4715-330 Braga, Portugal}
\affiliation{Departamento de F\'isica Aplicada, Universidad de Alicante, San Vicente del Raspeig, 03690 Spain}

\date{\today} 

\begin{abstract}
Quantum spin tunneling and Kondo effect are two very different  
quantum phenomena that produce the same effect on quantized spins, 
namely, the quenching of their magnetization. However, the nature of 
this quenching is very different so that quantum spin tunneling and Kondo effect compete 
with each other. 
Importantly, both quantum spin tunneling and Kondo effect produce very characteristic features 
in the spectral function that can be measured by means of single spin 
scanning tunneling spectroscopy and allows to probe the crossover 
from one regime to the other.  We model this crossover, and the 
resulting changes in transport, using a non-perturbative treatment of 
a generalized Anderson model including magnetic anisotropy that leads 
to quantum spin tunneling. We predict that, at zero magnetic field,  
integer spins can  feature a split-Kondo peak driven by quantum spin 
tunneling.
\end{abstract}

\maketitle

\section{Introduction}

Quantum spin tunneling (QST) and Kondo effect are two ubiquitous and widely studied \cite{Abragam-Bleaney,Gatteschi-book,Hewson-book} phenomena in the broad field of nanoscale magnetism. They both turn a spin system with a doubly degenerate ground state  into a system with a unique ground state with null magnetization. QST affects quantized integer spins with magnetic anisotropy, such as single molecule magnets \cite{Gatteschi-book}, magnetic impurities in insulators \cite{Abragam-Bleaney} and magnetic adatoms \cite{Hirjibehedin07,Khajetoorians2010} and molecules \cite{FePc} on surfaces.  The Kondo effect is most often associated with half-integer spins, but it has been observed in a variety of integer spin systems, such as quantum dots  with an even number of electrons \cite{Sasaki}, various integer spin magnetic molecules \cite{Parks2010,Tsukahara2011,Mugarza12} and molecular oxygen (spin $S=1$) adsorbed on gold \cite{Ho}.

The Kondo effect arises when a local spin is exchange coupled to itinerant electrons that respond dynamically to screen the magnetic moment of the impurity \cite{Hewson-book} This dynamical response leads to a resonance in the local density of states at zero energy, that emerges as a zero bias Fano feature  in the transport spectroscopy curves, $G(V)$, where $G\equiv \frac{dI}{dV}$. The observation of the Kondo resonance in individual magnetic atoms \cite{Madhavan98}  and molecules \cite{Tsukahara2011} by means of Scanning Tunneling Microscope (STM) spectroscopy has been reported numerous times in the last two decades. The natural energy scale that characterizes the Kondo effect is roughly given by the width of this resonance which  depends on the tunneling rate $\Gamma$ for electrons between the localized atomic orbitals and the extended states of the surface, and the charging energy of the atom, ${\cal U}_N\equiv E(N+1)-E(N)$.

QST can occur for integer spins with negative dominant 
uniaxial anisotropy $D<0$: In this case the ground state of a spin $S$ is doubly 
degenerate and consists of the two states with opposite and maximal spin projection 
$m_z=\pm{S}$, which are separated by an energy barrier $\sim{D}S^2$. A finite in-plane 
magnetic anisotropy $E$ then allows quantum tunneling between the two spin states,
lifting the degeneracy \cite{Garg93,Sessoli-Wernsdorfer} of the ground state by the tunnel splitting $\Delta_0$.
Thus the ground state is a linear combination of two spin states with opposite 
spin projection and null magnetization \cite{Delgado12,Delgado-2015}.
The dynamical spin response function acquires a pole at $\hbar\omega= \Delta_0$ so that QST can be probed spectroscopically.  
Whereas  in large spin molecular magnets $\Delta_0$ is so small that it can only be inferred indirectly \cite{Sessoli-Wernsdorfer},
for systems with  $S=1,2$, such as magnetic adatoms and small magnetic molecules deposited on conducting substrates, 
the spin excitations, and thereby the QST splittings, have been resolved directly by means of inelastic electron tunneling 
spectroscopy (IETS) using STM \cite{Hirjibehedin07}.
The corresponding $G(V)$ spectra show step features and, in general, no Kondo peak. It is customarily assumed that
these excitation energies are a property of the atomic/molecular quantized spin, weakly 
dressed \cite{Hirjibehedin07,Oberg14,Delgado-Surf-Sci,Delgado-2015} by its Kondo exchange coupling. 

Whether a given magnetic atom or molecule  will show a stepwise $G(V)$ spectrum \cite{Hirjibehedin07,FePc,Khajetoorians2010} or a zero bias Kondo resonance depends on the strength of the Kondo exchange $J$ which is controlled by the ratio $\Gamma/{\cal U}$. Thus, FePc molecules, that in vacuum have $S=1$, display a Kondo feature when deposited on Au(111)\cite{Tsukahara2011} and  inelastic steps when deposited on oxidized Cu(110) \cite{FePc}. Moreover, the joint observation of a zero bias Kondo resonance together with stepwise inelastic spin excitations has been reported for individual magnetic atoms \cite{Otte08,Oberg14}. Importantly, it is possible to devise experiments \cite{Parks2010,Oberg14,Jacobson15} in which the Kondo interaction could be tuned, making it relevant to address the question of how the $G(V)$ spectra evolve from the weak to the strong coupling regime.  
The magnitude  of the quantum spin tunneling splitting can also be modulated by application of a magnetic field along the hard axis direction \cite{Garg93,Sessoli-Wernsdorfer}.
 In this work  we address how the competition between Kondo screening and QST affect the STM inelastic conductance  and we predict a new physical phenomenon, the splitting of the Kondo peak at zero magnetic field due to quantum spin tunneling.

Both the inelastic steps \cite{JFR09} and the Kondo features \cite{Zitko,Sanvito,Ternes} can be described using a 
Kondo Hamiltonian where the atomic spin is described with a single-ion quantized  spin interacting, 
via exchange, with the conduction electrons of the surface. In particular, the interplay between 
magnetic anisotropy and Kondo screening has been thoroughly studied using the Kondo 
model \cite{Zitko,Sanvito,Oberg14,Delgado-Surf-Sci,Ternes}.
Here we present a more general approach, based on  a non-perturbative treatment of
a multi-orbital Anderson model for the adatom coupled to the substrate, that permits to 
include atomic charge fluctuations that are effectively frozen in the Kondo model. 
As we show below, valence fluctuations have a similar effect on the spectra as Kondo exchange.
In addition, density functional theory (DFT) calculations show that often charge is not quantized in
magnetic adatom systems \cite{Ferron2015,Panda2015}.

\begin{figure}
  \begin{center}
    \begin{tabular*}{\linewidth}{@{\extracolsep{\fill}}cc}
      \includegraphics[width=0.35\linewidth]{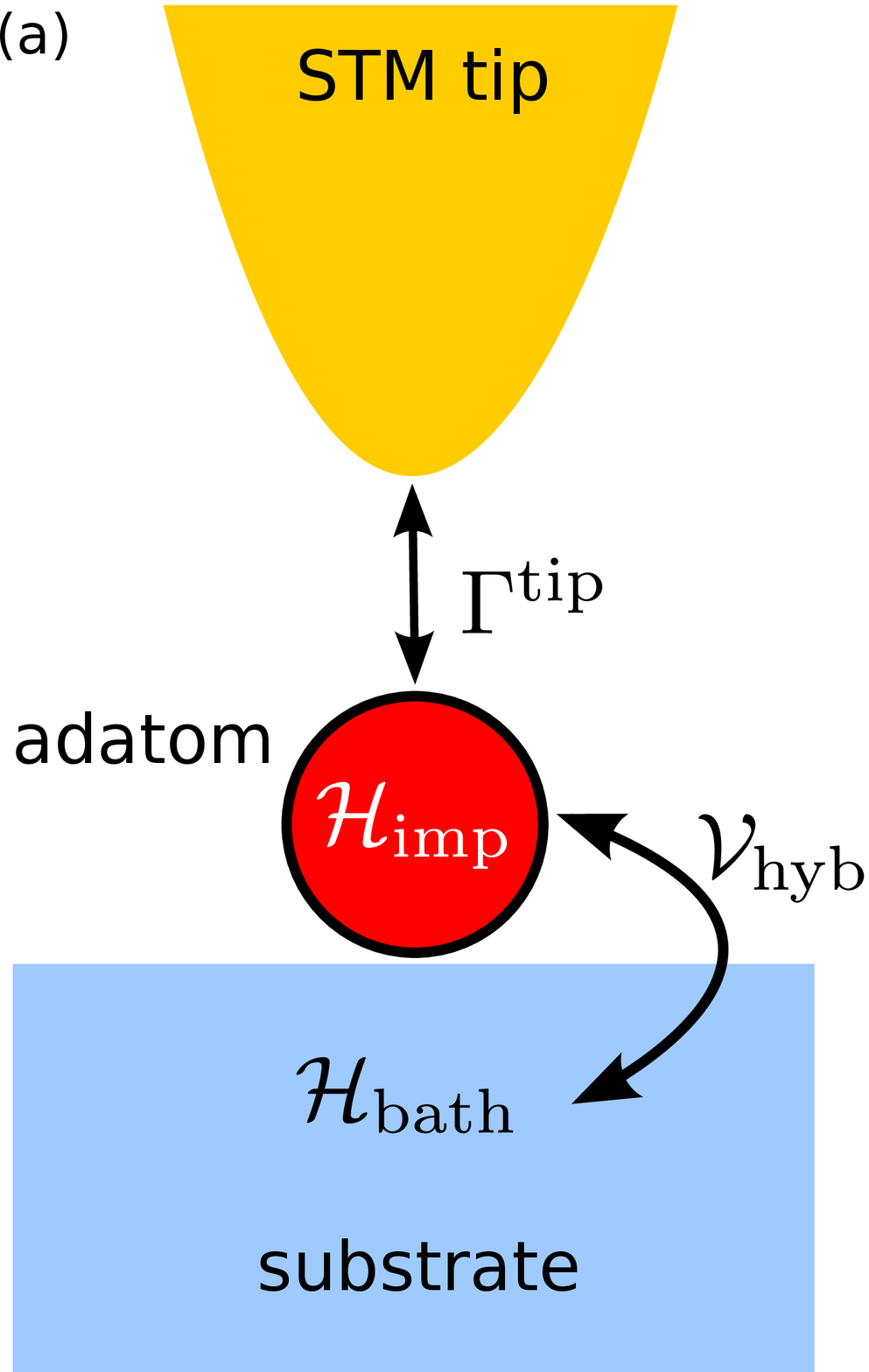} & 
      \includegraphics[width=0.55\linewidth]{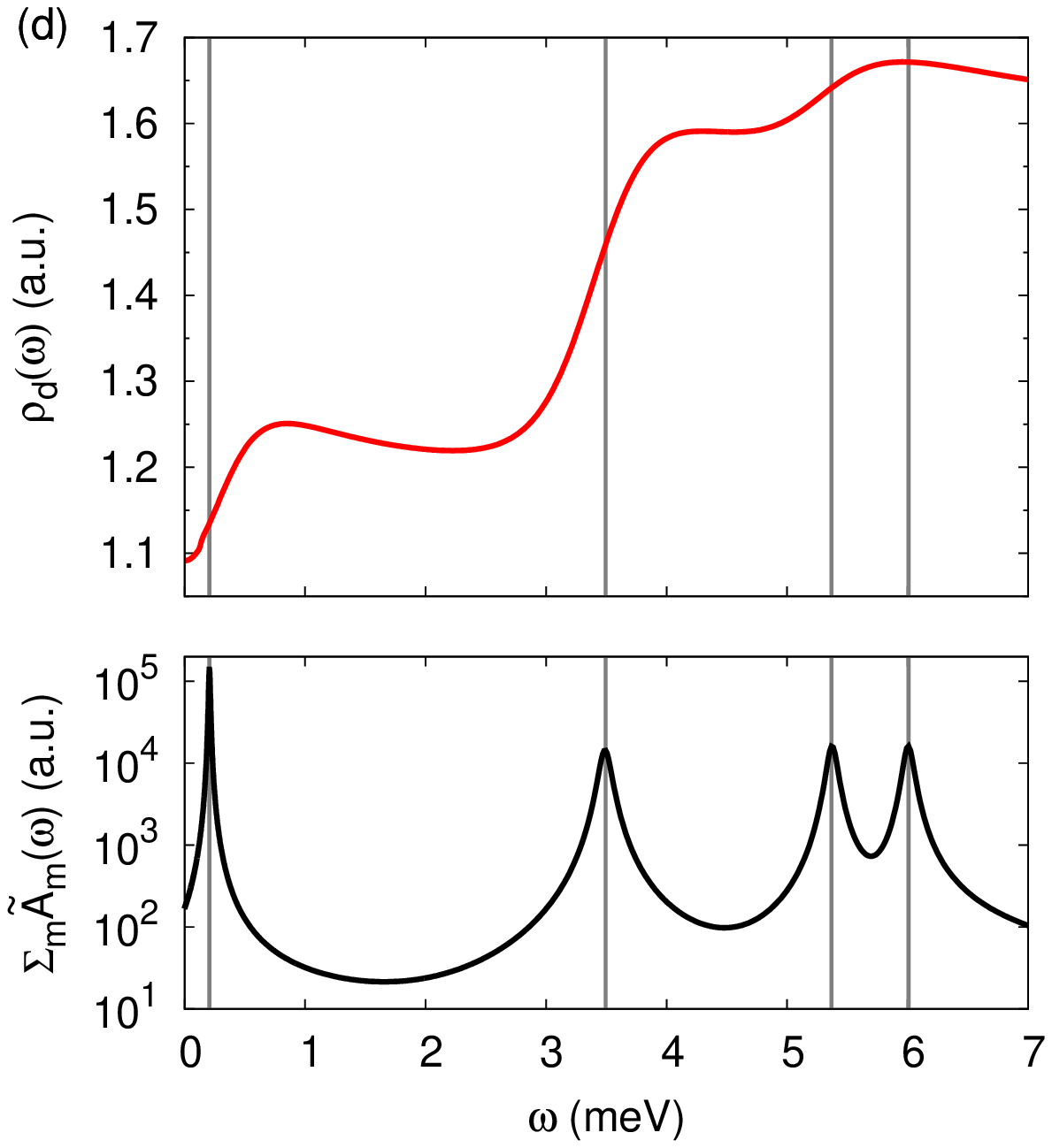} \\
      \includegraphics[width=0.4286\linewidth]{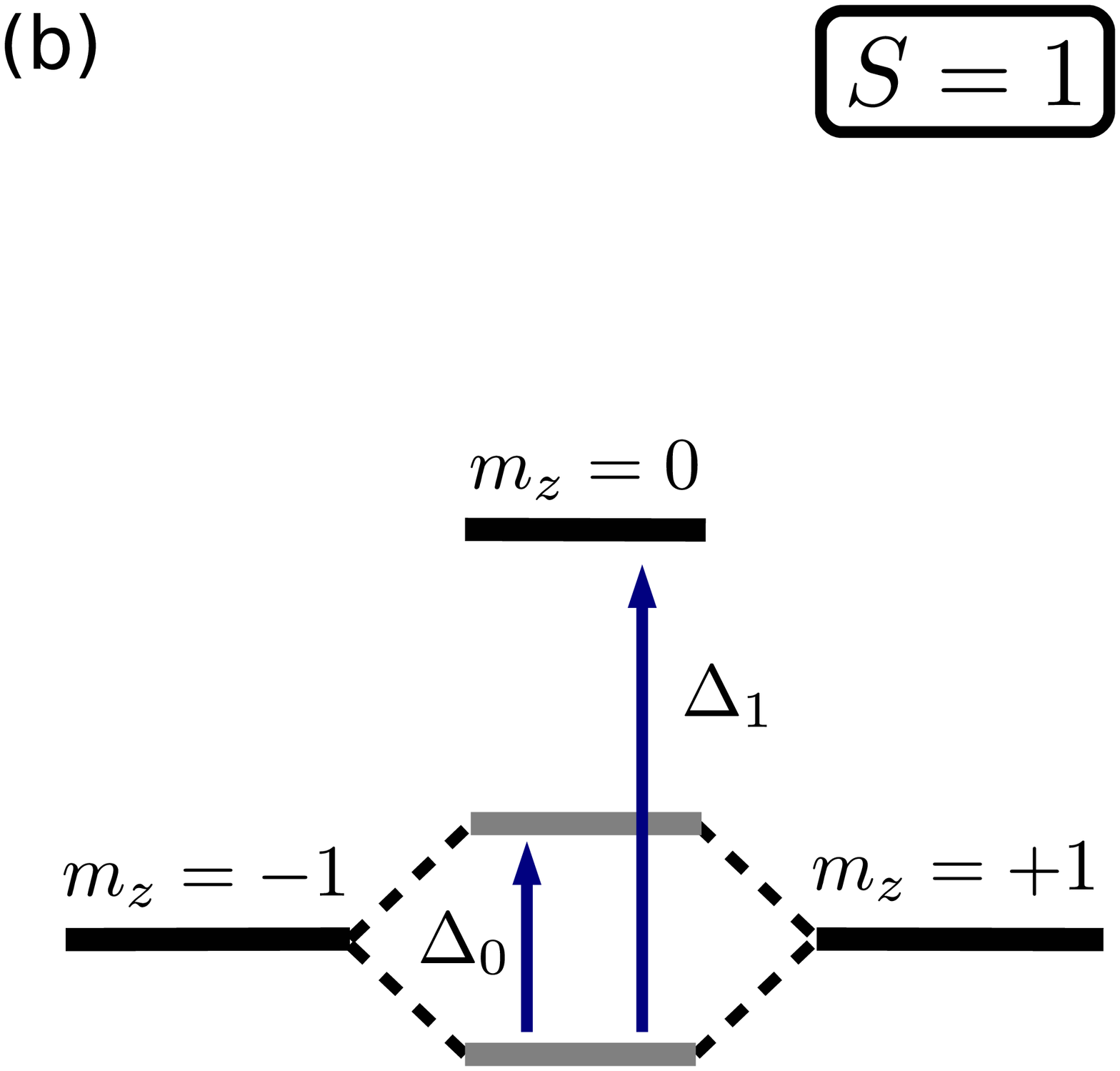} & 
      \includegraphics[width=0.5714\linewidth]{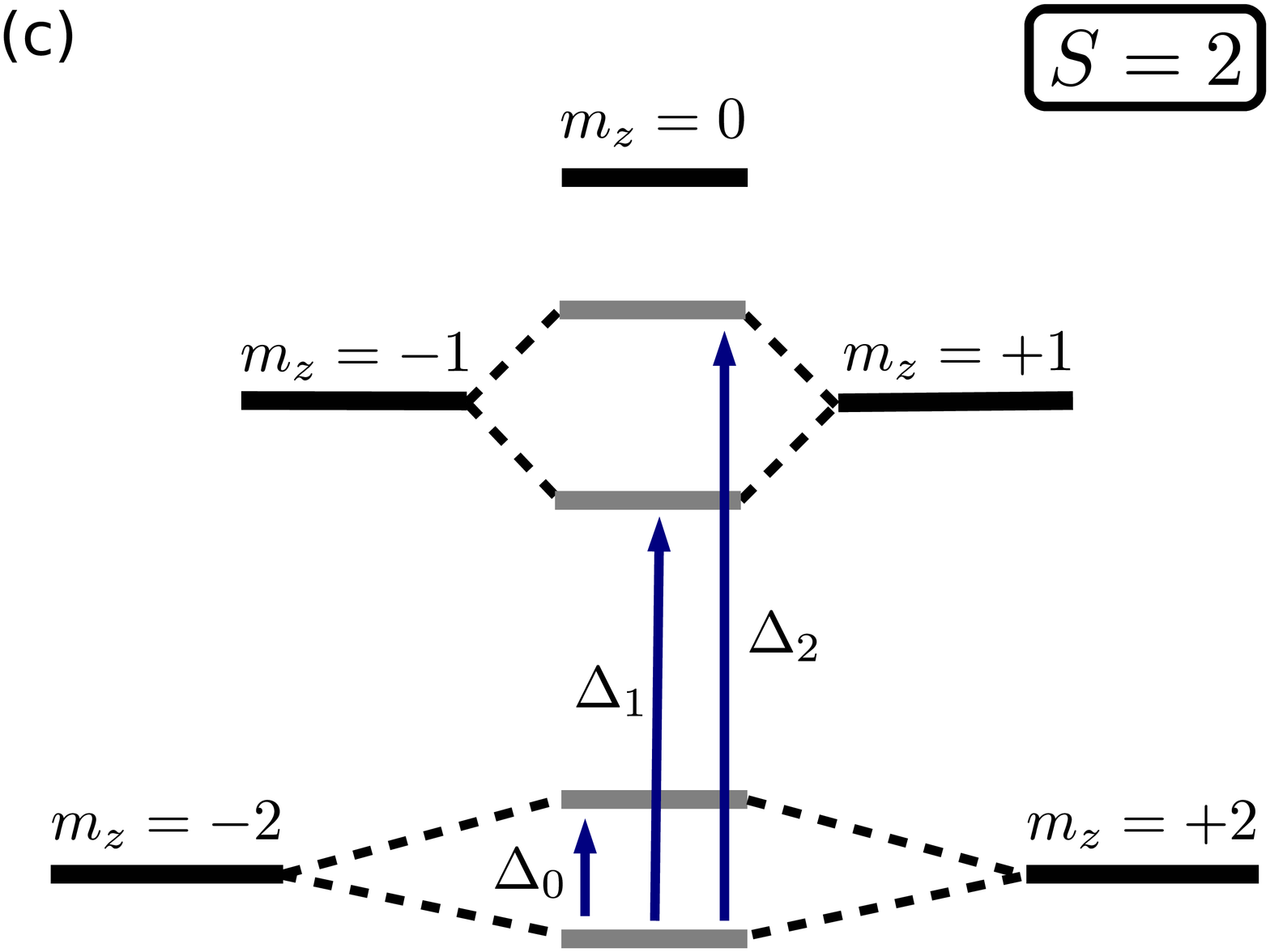}       
    \end{tabular*}
  \end{center}
  \caption{
    \label{fig:method}
    (a) Schematic model of experimental setup for measuring
    excitation spectra of a magnetic adatom on a surface
    with STM tip.
    (b) Schematic energy spectrum of adatom GS multiplet 
    for $S=1$ for negative uniaxial anisotropy 
    $D<0$ and finite in-plane anisotropy $E$.
    (c) Same as (b) but for $S=2$.
    (d) Comparison of PP spectrum $\tilde{A}_m(\omega)=A_m(\omega)/f(-\omega)$
    for the four lowest excitations (bottom) and the real 
    electron spectrum (top) for $S=2$, $D=-1.55meV$ and $E=0.35meV$.
    The vertical grey lines indicate the positions of the 
    PP energies $E_m^\ast$ w.r.t. the GS PP energy $E_0^\ast$.
  }
\end{figure}

\section{Model and method}

We consider tunneling between an STM tip and a magnetic adatom coupled to a surface
as shown in Fig.~\ref{fig:method}(a).
Assuming weak coupling to the STM tip (tunneling regime) \cite{Datta-book}  
the low-bias conductance can be directly related to the 
adatom density of states (technically, the many-body spectral function) $\rho_\alpha(\omega)$,  
\begin{equation}
  \mathcal{G}(V) = \frac{2e^2}{\hbar} \sum_{\alpha} \Gamma^{\rm tip}_{\alpha} \rho_{\alpha}(eV) 
  \label{GdeV}
\end{equation}
where $\Gamma^{\rm tip}_{\alpha}=\pi|{V_{\alpha}^{\rm tip}}|^2\rho_{\rm tip}$ 
is the (weak) tunneling rate of electrons between the adatom orbitals $\alpha$ and  the STM tip.
We have assumed the DOS of the STM tip $\rho_{\rm tip}$ is energy independent around the 
Fermi level. 
Note however, that in general different orbitals couple differently 
to the STM tip so that the contribution of the individual channels 
to the total conductance may differ.
We neglect direct tunneling into surface states  in eq.~(\ref{GdeV}). This is 
a good approximation when the magnetic atoms are separated from the metallic surface
by a decoupling insulating layer, such as Cu$_2$N/Cu(100) \cite{Hirjibehedin07,Oberg14,Otte08},
CuO/Cu \cite{FePc} and h-BN/Rh(111) \cite{Jacobson15}. This approximation does not
capture the Fano interference effect relevant \cite{Madhavan98} when the tip-atom 
channel interferes with the direct tip-surface tunneling path \cite{Ujsaghy2000}.

We describe the magnetic atom on the surface  by a multi-orbital Anderson model,
\begin{equation}
  {\cal H} = \mathcal{H}_{\rm imp} + {\mathcal H}_{\rm bath} + {\cal V}_{\rm hyb}
  \label{eq:AIM}
\end{equation}
where the Hamiltonian of the Anderson impurity site $\mathcal{H}_{\rm imp}$ 
describes the strongly interacting $3d$-levels that yield the spin of 
the magnetic atom, and includes a term that accounts for magnetic anisotropy:
\begin{eqnarray}
  {\cal H}_{\rm imp} &=& \epsilon_d \hat{N}_d 
  + \sum_{\alpha\sigma\neq\alpha^\prime\sigma^\prime} U \, \hat{n}_{\alpha\sigma} \, \hat{n}_{\alpha^\prime\sigma^\prime} 
  - \sum_{\alpha\neq{\alpha^\prime}} J_{\rm H} \, \vec{S}_\alpha \cdot \vec{S}_{\alpha^\prime} 
  \nonumber\\
  &+& D \hat{S}_z^2 + E (\hat{S}_x^2 - \hat{S}_y^2 )
  \label{eq:imp}
\end{eqnarray}
$\epsilon_d$ are the single-particle energies of the $d$-levels,
$\hat{N}_d=\sum_{\alpha,\sigma}\hat{n}_{\alpha\sigma}$ is the number operator
for all $d$-levels $\alpha=1,\ldots,M$, $\hat{n}_{\alpha\sigma}=d_{\alpha\sigma}^\dagger d_{\alpha\sigma}$ 
is the number operator of an individual $d$-level $\alpha$ with spin $\sigma$, 
$U$ is the effective Coulomb repulsion, $J_{\rm H}$ the Hund's coupling, 
and $\vec{S}_\alpha$ measures the total spin of an individual $d$-level $\alpha$, 
i.e. $\vec{S}_\alpha= \sum_{\sigma\sigma^\prime} d_{\alpha\sigma}^\dagger \vec\tau_{\sigma\sigma^\prime} d_{\alpha\sigma^\prime}$.
The crystal field splitting of the $d$-levels 
together with the spin-orbit coupling (SOC) gives rise \cite{Abragam-Bleaney} 
to magnetic anisotropy (MA) which in our simplified model is taken into account by the effective
spin Hamiltonian given by the last term of (\ref{eq:imp}) where $D$ is the
uniaxial anisotropy and $E$ the in-plane anisotropy \cite{Gatteschi-book}.

The second term in (\ref{eq:AIM}) describes the conduction electron bath in the surface:
\begin{equation}
  {\cal H}_{\rm bath}=\sum_{k,\alpha,\sigma} \varepsilon_{k\alpha} c_{k\alpha\sigma}^{\dagger}c_{k\alpha\sigma}
\end{equation}
The third term in (\ref{eq:AIM}) is the so-called hybridization term which describes the 
coupling between the impurity and the conduction electron bath:
\begin{equation}
  \label{eq:Vhyb}
  {\cal V}_{\rm hyb} = \sum_{k,\alpha,\sigma} V_{k\alpha} (c_{k\alpha\sigma}^\dagger d_{\alpha\sigma} + d_{\alpha\sigma}^\dagger c_{k\alpha\sigma})
\end{equation}
Integrating out the bath degrees of freedom one obtains the so-called hybridization function:
\begin{equation}
  \label{eq:hybfunc}
  \Delta_\alpha^{\rm hyb}(\omega)=\sum_{k}\frac{|V_{k\alpha}|^2}{\omega+\mu-\varepsilon_{k\alpha}+i\eta}
\end{equation}
Its (negative) imaginary part $\Gamma_\alpha(\omega)=-\Im\,\Delta_\alpha^{\rm hyb}(\omega)$ describes the single-particle broadening 
of individual impurity levels $\alpha$ due to the coupling to the conduction electrons.
Note that we have assumed here that each impurity level $\alpha$ couples to different conduction electron states 
(labeled by $\alpha$) in the substrate so that each impurity level has its own bath. This assumption is justified 
because of the different symmetries of the $d$-orbitals. In the case of coupling of two orbitals to the same conduction
electron states, off-diagonal elements in the hybridization function $\Delta_{\alpha\alpha^\prime}^{\rm hyb}(\omega)$ would occur,
describing substrate mediated hopping between impurity levels. In adatom-substrate systems these off-diagonal elements 
are often either zero or very small \cite{adatom_hyb}.

We now solve the Anderson model (2) within the so-called One-Crossing Approximation (OCA) \cite{Pruschke,Haule:2001,Haule:2010}.
The first step is an exact diagonalization of the \emph{isolated} impurity Hamiltonian (\ref{eq:imp}):
\begin{equation}
  \label{eq:Himp_diag}
  {\cal H}_{\rm imp} = \sum_m E_m \ket{m}\bra{m}
\end{equation}
The many-body eigenstates $\ket{m}$ are eigenstates of the total number of electrons of the impurity, 
i.e. $\hat{N}_d\ket{m}=N_m\ket{m}$, and the total spin $S^2$ of the impurity, i.e. $\hat{S}^2\ket{m}=S_m(S_m+1)\ket{m}$.
For a ground state (GS) with integer spin $S$, and  magnetic anisotropy with negative uniaxial
anisotropy $D\ne0$ and finite in-plane anisotropy $E\ne0$, 
the  $(2S+1)$ degeneracy of the GS multiplet is completely lifted. This is schematically
shown on the right hand side of Fig.~\ref{fig:method} for $S=1$ and $S=2$ and $D<0$ \cite{Abragam-Bleaney,Hirjibehedin07}:
For $D<0$ and $E=0$ the GS is doubly degenerate with the GS doublet having the maximal spin 
projection $m_z=\pm{S}$. A finite in-plane anisotropy $E$ allows for \emph{quantum tunneling}
between both spin directions thus lifting the degeneracy of the GS doublet, which now becomes 
split by $\Delta_0$, the  bare quantum spin tunneling.
The quantum states of the split doublet are thus linear combinations $\ket{\pm}\sim\ket{m_z=+S}\pm\ket{m_z=-S}$.

In the next step a diagrammatic expansion of the many-body eigenstates $\ket{m}$ 
of the (isolated) impurity ${\cal H}_{\rm imp}$ in terms of the hybridization
(\ref{eq:Vhyb}) is developed. ${\cal V}_{\rm hyb}$ connects eigenstates $\ket{m}$ 
and $\ket{n}$ of ${\cal H}_{\rm imp}$ with occupation numbers differing by one 
($N_m=N_n\pm1$, see (\ref{eq:Vhyb_mb}) in App.~\ref{App:OCA}).
It is these fluctuations between the impurity GS and excited states with one more or 
one less electron that give rise to the Kondo effect.  
To this end one introduces so-called pseudo-particles (PPs) $m$ corresponding to 
the many-body eigenstates $\ket{m}$. The full propagator of such a PP $m$ can be 
written as
\begin{equation}
  \label{eq:PPGF}
  G_m(\omega) = \frac{1}{\omega-\lambda-E_m-\Sigma_m(\omega)}
\end{equation}
where $\Sigma_m(\omega)$ is the PP self-energy which describes the 
renormalization (real part) and broadening (imaginary part) of the 
PP $m$ due to the interaction with other PPs $m^\prime$ mediated by 
the conduction electron bath [see eq. (\ref{eq:pp_interaction}) in App. \ref{App:OCA}].
$-\lambda$ is the chemical potential for the PPs which has to be adjusted
such that the total PP charge is constrained to one [see 
eqs.~(\ref{eq:pp_charge},\ref{eq:Himp_pp}) in App.~\ref{App:OCA}].

OCA consists in a diagrammatic expansion of the PP self-energies $\Sigma_m$
in terms of the hybridization function $\Delta_\alpha^{\rm hyb}$ to 
infinite order but summing only a subset of diagrams (only those involving 
conduction electron lines crossing at most once).
This leads to a set of coupled integral equations for the PP propagators
and self-energies that have to be solved self-consistently. 
Once the OCA equations are solved the real electron spectral function 
$\rho_\alpha(\omega)$ for the impurity levels entering equation 
(\ref{GdeV}) for calculating the conductance spectrum
is obtained from convolutions of the PP spectral 
functions $A_m(\omega)=-\Im\,G_m(\omega)/\pi$ [see eqs. (\ref{eq:real_gf}-\ref{eq:rho_d}) in App.~\ref{App:OCA}] which feature sharp
resonances at the renormalized many-body energies $E_m^\ast=E_m+\Re\,\Sigma_m(E_m^\ast)$.
The differences between the renormalized energies of the excited states $E_m^\ast$ 
and the GS $E_0^\ast$ yield the real electronic excitations as shown Fig.~1(d).

OCA captures both the weak and strong coupling Kondo regimes, and has shown 
to produce reliable spectra for the single-orbital Anderson model, as 
long as the temperatures are not too low compared to the Kondo 
temperature \cite{OCA-artifacts}.
For the general multi-orbital situation considered here, benchmarking of spectra 
is difficult since Numerical Renormalization Group \cite{NRG} is computationally too demanding 
to be applied. Using Continuous-Time Quantum Monte-Carlo, it has been shown \cite{Ruegg13}
that  OCA is markedly superior to the simpler Non-Crossing Approximation (NCA) 
for dealing with multi-orbital Anderson models, although 
certain sum rule violations are  found.
Very importantly, OCA has shown excellent agreement with experiments
in very complex multi-orbital systems \cite{Jacob13,Oberg14,Karan15}.
More details about the OCA method are given in App.~\ref{App:OCA}.

\section{Results}

\begin{figure}
  \begin{tabular}{cc}
    \includegraphics[width=0.49\linewidth]{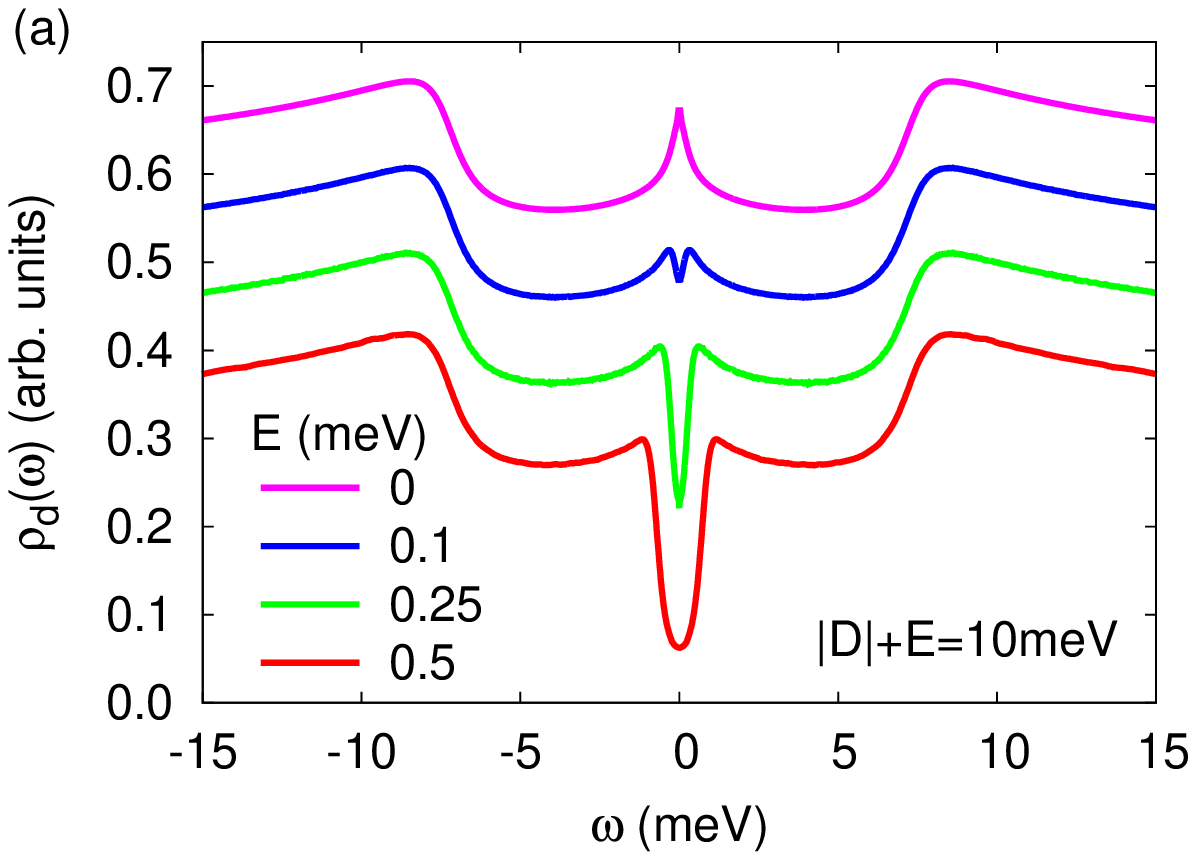} &
    \includegraphics[width=0.49\linewidth]{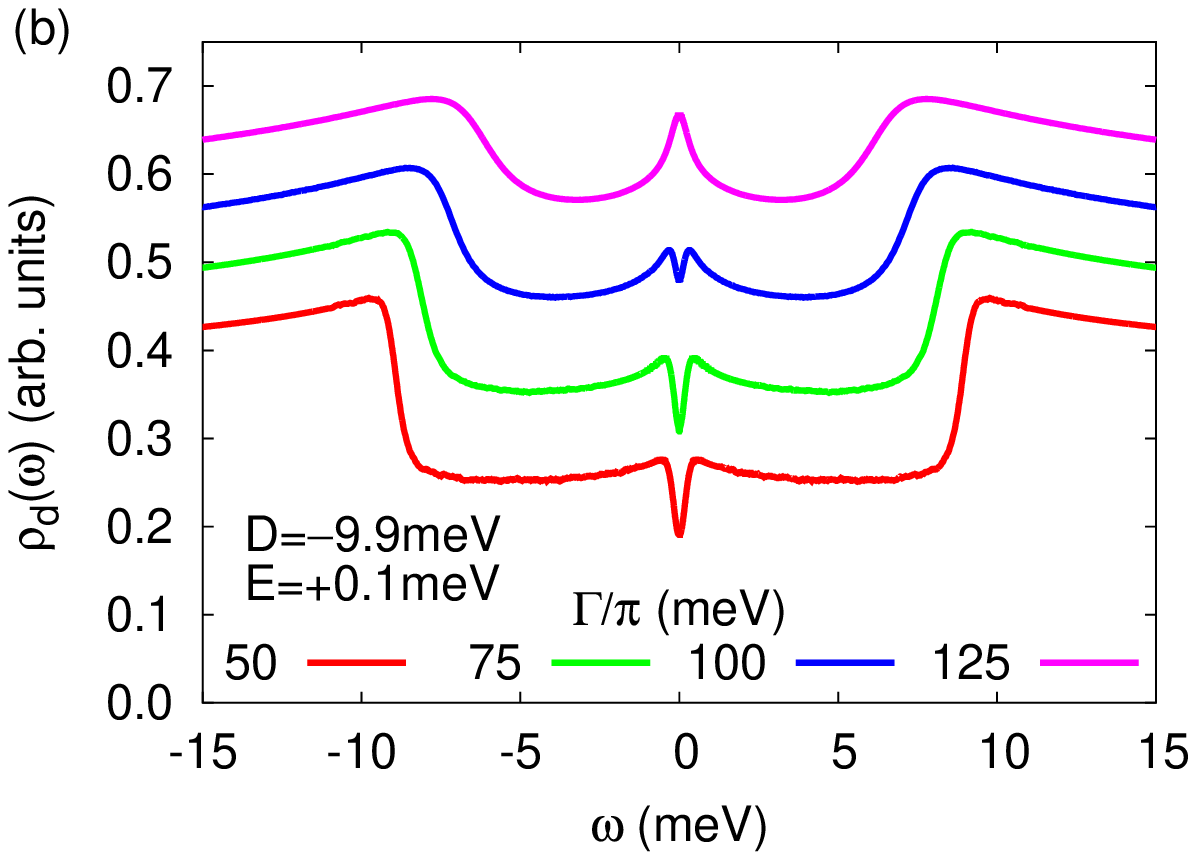} \\
    \includegraphics[width=0.49\linewidth]{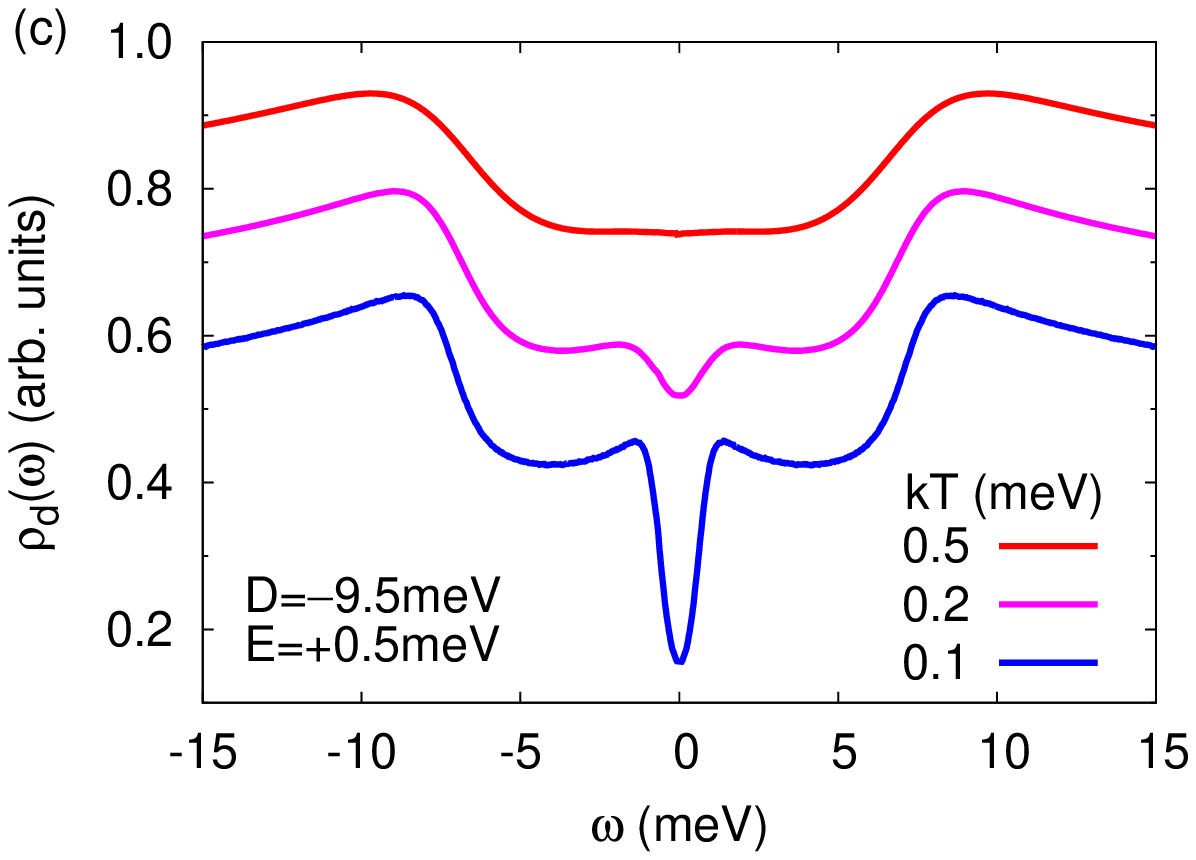} &
    \includegraphics[width=0.49\linewidth]{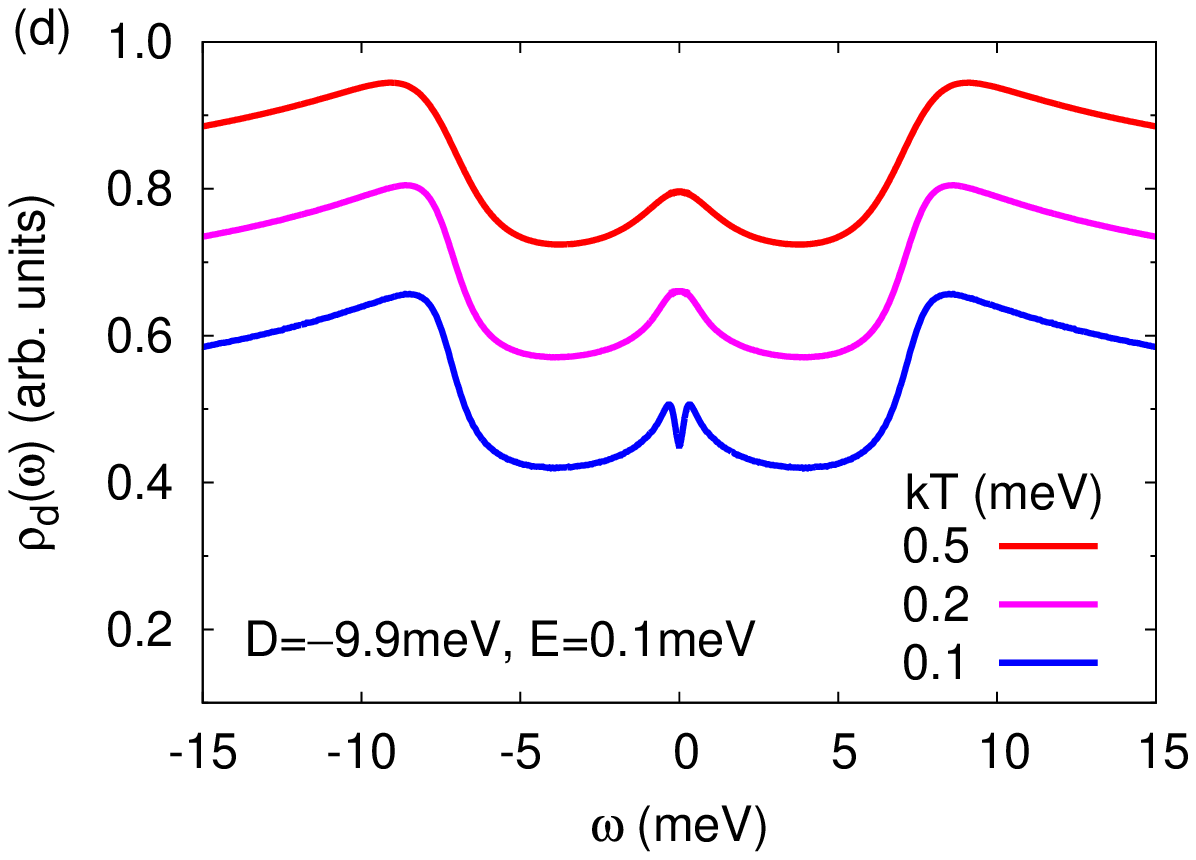} \\
    \includegraphics[width=0.49\linewidth]{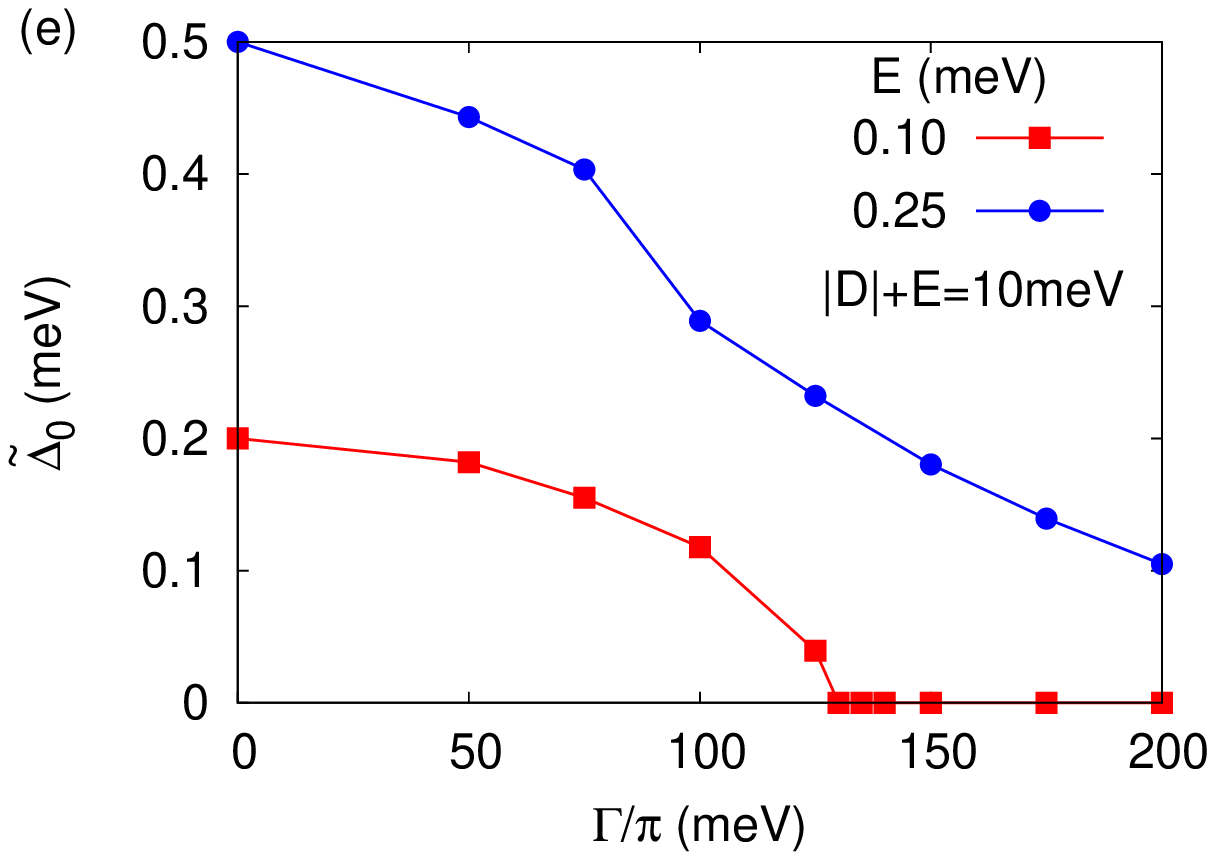} &
    \includegraphics[width=0.49\linewidth]{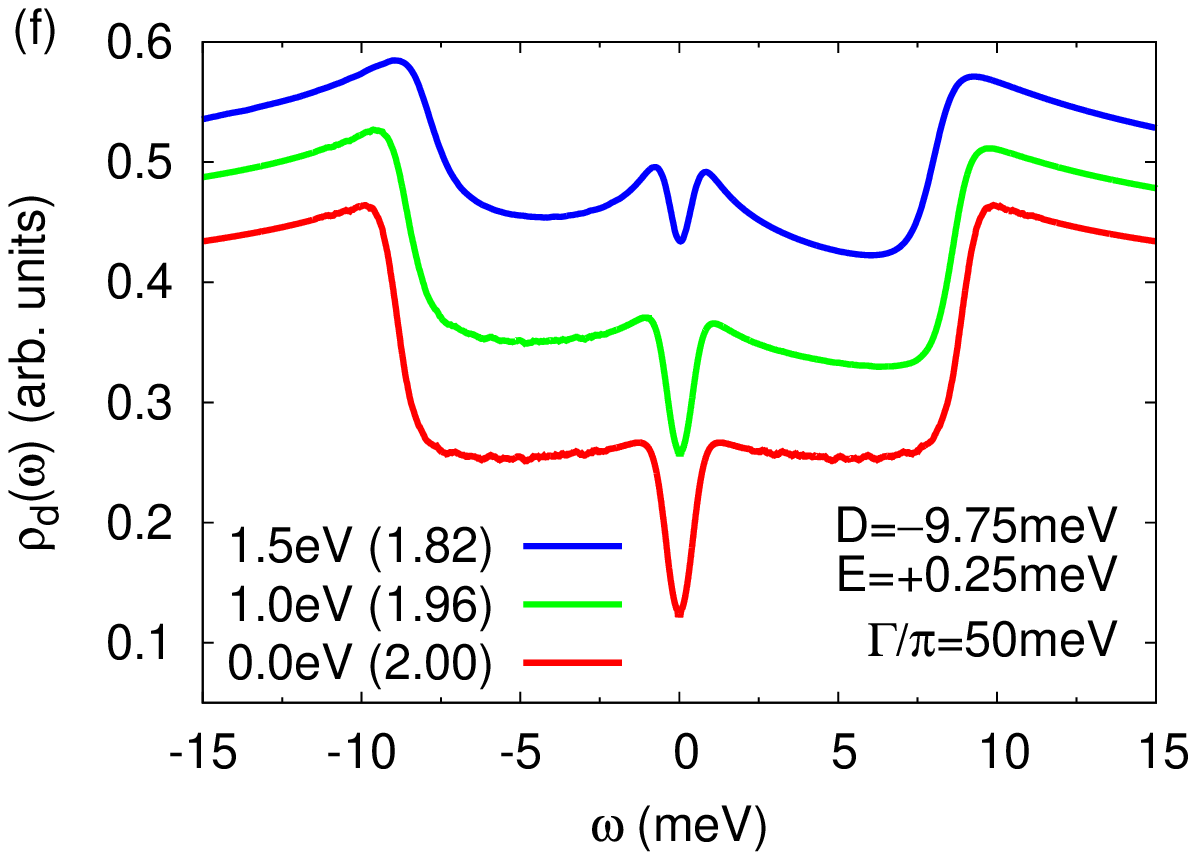}     
  \end{tabular}
  \caption{
    \label{fig:spin-1}
    Results for spin $S=1$ and for $D<0$, and temperature $kT=0.1$meV.
    (a-d) Spectra $\rho_d(\omega)$ in dependence of different system
    parameters: (a) Changing $E$ while keeping fixed $\Delta_1=|D|+E=$10meV and $\Gamma/\pi=100$meV.
    (b) Changing $\Gamma$ for $D=-9.9$meV and $E=0.1$meV.
    (c) Temperature dependence for $\Gamma/\pi=100$meV, $D=-9.75$meV and $E=0.25$meV. 
    (d) Same as (c) but for $D=-9.9$meV and $E=0.1$meV.
    (e) Effective QST $\tilde\Delta_0$ as a function of $\Gamma$.
    (f) Spectra for $D=-9.75$meV, $E=0.25$meV and $\Gamma/\pi=100$meV 
    for different energy level shifts $\delta\epsilon_d$.  
    In parenthesis the total occupancy $N_d$ of the impurity is given.  
  }
\end{figure}

We assume that the spin $S$ of the impurity is carried by $2S$ active impurity levels 
close to half-filling. For all impurity levels we assume the same energy-independent 
broadening $\Gamma$. 
The important energy scale is the Kondo exchange coupling $J\sim\Gamma/{\cal U}$ 
which can be varied by either changing $\Gamma$ or the charging energy ${\cal U}$
which depends on the Coulomb repulsion $U$.
Here we choose to vary $\Gamma$ and fix the Coulomb repulsion to $U=4$eV.
The Hund's coupling whose main effect is to favor the formation 
of a high-spin ground state is fixed to $J_{\rm H}=1$eV.

We first study the competition between Kondo quenching and QST for a 
magnetic atom with  spin $S=1$, at the point of electron-hole (e-h) symmetry.
In this case the competition is controlled 
by two energy scales, the in-plane anisotropy $E$, that drives the QST, and $\Gamma$, 
that favors Kondo coupling. We assume $D<0$ so that, for $E=0$, the ground state is 
the doublet of eigenstates of $S_z$ with $m_z=\pm1$ and the state with $m_z=0$ is the 
first excited state [see Fig.~\ref{fig:method}(b)]. 
Because of the two-fold degeneracy of the GS
the Kondo effect can take place [top curve in Fig.~\ref{fig:spin-1}(a)]. 
Spin-flip events for the GS doublet $m=\pm1$ occur via the excited state $m=0$ and hence are reduced by a factor of 
$J/D$ compared to the Kondo exchange $J$ of the corresponding spin-1 Kondo model
without anisotropy.
Note that such a spin-flip process involves the simultaneous 
scattering of two conduction electrons. This is possible since we are considering 
a multichannel situation as each impurity orbital is connected to its own bath.
In the situation of just a single screening channel, the spin-flip between the 
$m=\pm1$ GS doublet would be inhibited when $D$ becomes bigger than the Kondo 
temperature, leading to a split-Kondo feature \cite{Romeike:2006,Zyazin:2010}.

The in-plane anisotropy term $E>0$ produces the QST that leads to a splitting of 
the $m=\pm1$ doublet, so that the bare excitation energies are $\Delta_0=2E$ 
and $\Delta_1=|D|+E$. At weak coupling (small $\Gamma$), the spectra show two 
steps corresponding to inelastic spin transitions between the renormalized spin 
levels [bottom curve in Fig.~\ref{fig:spin-1}(a)].  
Our calculations show that the effect of decreasing $E$ [Fig.~\ref{fig:spin-1}(a)] and 
increasing $\Gamma$ [Fig.~\ref{fig:spin-1}(b)] is similar. In both instances the renormalized 
QST splitting $\tilde\Delta_0$ decreases, and the line shapes evolve from square steps 
at small $\Gamma$ or large $E$ to a characteristic triangular shape, very often seen in 
experiments \cite{Hirjibehedin07,Otte08,Khajetoorians2010}, that can only be captured in part when going 
beyond second order perturbation theory in the Kondo exchange \cite{Sanvito,Ternes}.
Whereas the reduction of the renormalized QST splitting $\tilde\Delta_0$ as $E$ decreases 
is trivially accounted for by the fact that $\Delta_0=2E$, the red-shift renormalization 
of the spin excitation energies due to Kondo coupling [Fig.~\ref{fig:spin-1}(e)] is a 
many-body effect, in line with previous results \cite{Zitko11,Korytar12,Oberg14,Delgado-Surf-Sci,Delgado-2015}.
In our approach the renormalization of the spin excitations ultimately 
originates in the renormalization of the many-body energies $E_m^\ast$ by the real part of 
the PP self-energy $\Sigma_m(\omega)$, see eq. (\ref{eq:PPGF}).
We would like to stress at this point that the splitting of the Kondo peak 
is exactly given by the effective QST $\tilde\Delta_0$. Hence Fig.~\ref{fig:spin-1}(e) 
really is a prediction for the splitting of the Kondo peak by QST which can be measured
experimentally.
 
At intermediate coupling, the renormalization of the line shape and energy of the lowest energy 
spin excitation turns them into a split Kondo peak. This is one of the important results of this 
work: In the absence of a magnetic field the competition between QST and Kondo effect yields a 
split-Kondo state whose characteristic signature can be measured by STM spectroscopy. 
As shown in Figs.~\ref{fig:spin-1}(c,d), the temperature dependence of the spectra 
in the case of intermediate coupling (split-Kondo) is quite different from that of weak 
coupling (step-like).
For weak coupling [Fig.~\ref{fig:spin-1}(c)],  at low temperatures two clear  steps are obtained, with
slight triangular departures from the step-wise behavior,  
due to Kondo interactions, that are  captured as well by perturbative calculations \cite{Ternes}. As
the temperature is increased, the low energy step is smeared out and, eventually, is no longer resolved.
On the other hand, at intermediate coupling [Fig.~\ref{fig:spin-1}(d)], a zero-field split Kondo peak 
is obtained at low temperatures. As we increase $k_BT$
the QST splitting of the Kondo peak disappears  due to thermal smearing, 
resulting in a single peak.
At strong coupling, the effective QST $\tilde\Delta_0$ vanishes [Fig.~\ref{fig:spin-1}(e)], 
so that the ground state and the first excited state become effectively degenerate: 
Kondo coupling quenches QST splitting \cite{Delgado-2015}.
At this point, the spectra show a Kondo peak [top curves in Figs.~\ref{fig:spin-1}(a,b)].

We now consider the effect of valence fluctuations. This effect is particularly important since 
DFT calculations show that often the average occupation of the $d$ shell is not quantized \cite{Ferron2015,Panda2015}.
In our model, we control the valence mixing by shifting 
the impurity levels by an amount $\delta\epsilon_d$, taking the system out of the e-h symmetry point, 
leading to deviations from integer occupation number of the impurity shell.
Valence fluctuations have a similar effect on the spectra as increasing the Kondo screening, 
leading e.g. to a considerable enhancement of the Kondo peak \cite{valence-fluctuations}.
Similarly, also the renormalization of the excitation energies and especially of the QST 
by the Kondo is enhanced by the introduction of charge fluctuations, as can be seen in 
Fig.~\ref{fig:spin-1}(f).
Increasing $\delta\epsilon_d$ leads to the step features associated with spin excitations moving to 
lower energies. Also the up-bending of the lowest energy steps associated with QST and the conversion 
into a split-Kondo peak is induced by increasing valence fluctuations. Different from the case when 
$\Gamma$ increases, the spectra also become somewhat asymmetric as the e-h symmetry is broken 
by the valence fluctuations. 

\begin{figure}
  \begin{tabular}{cc}
    \includegraphics[width=0.49\linewidth]{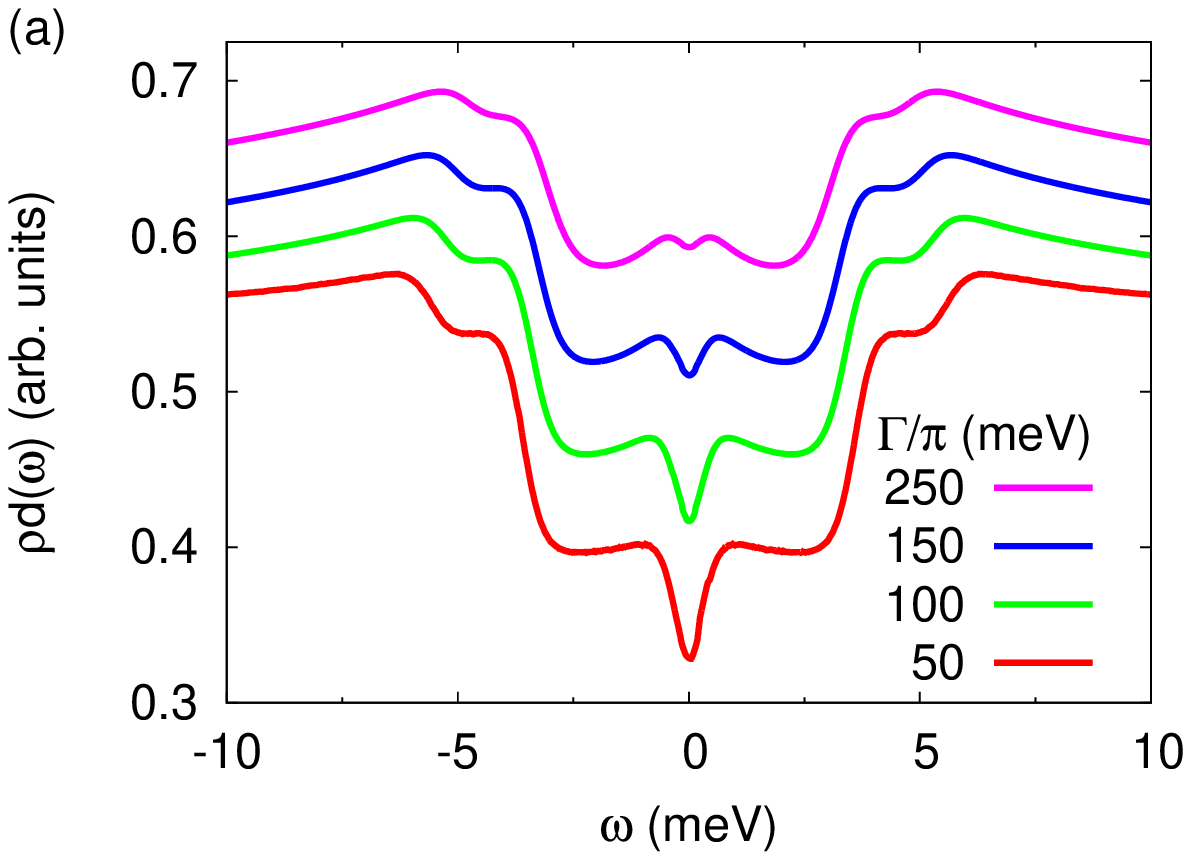} & 
    \includegraphics[width=0.49\linewidth]{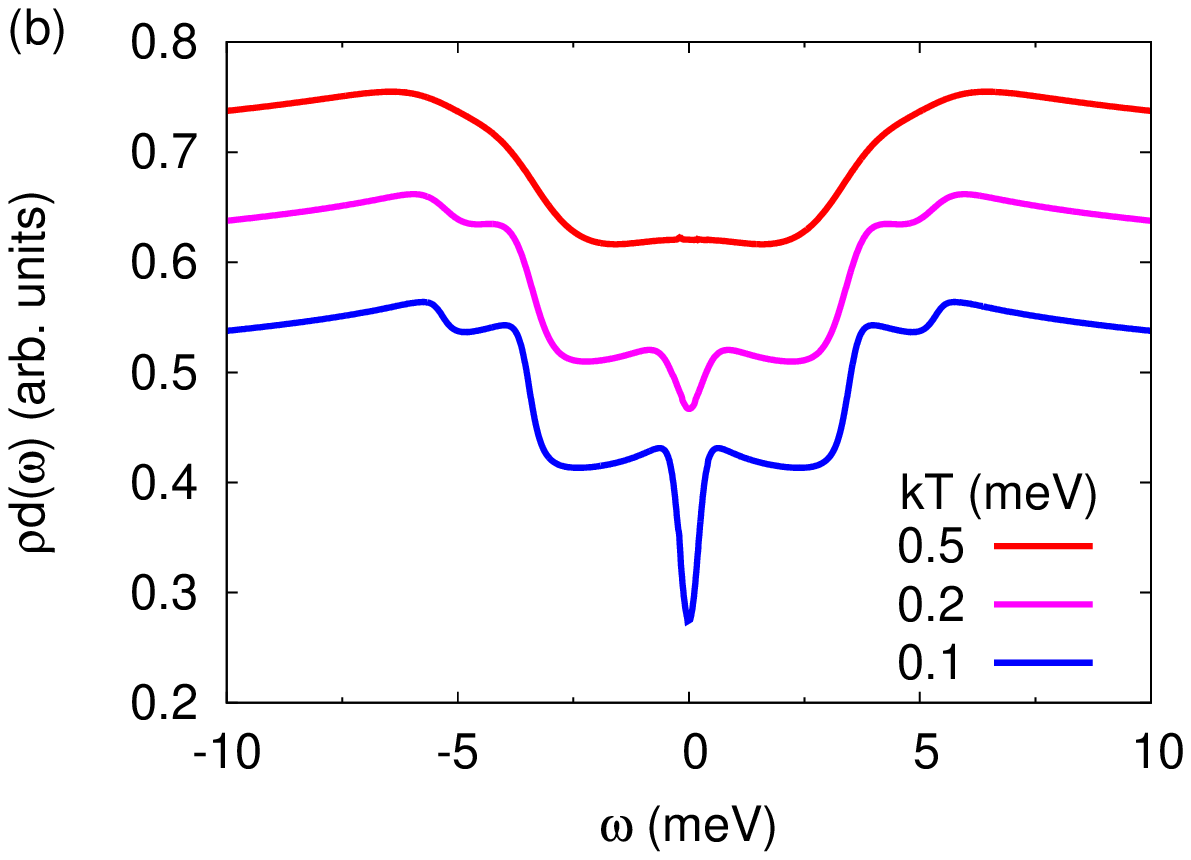} 
    \\
    \includegraphics[width=0.49\linewidth]{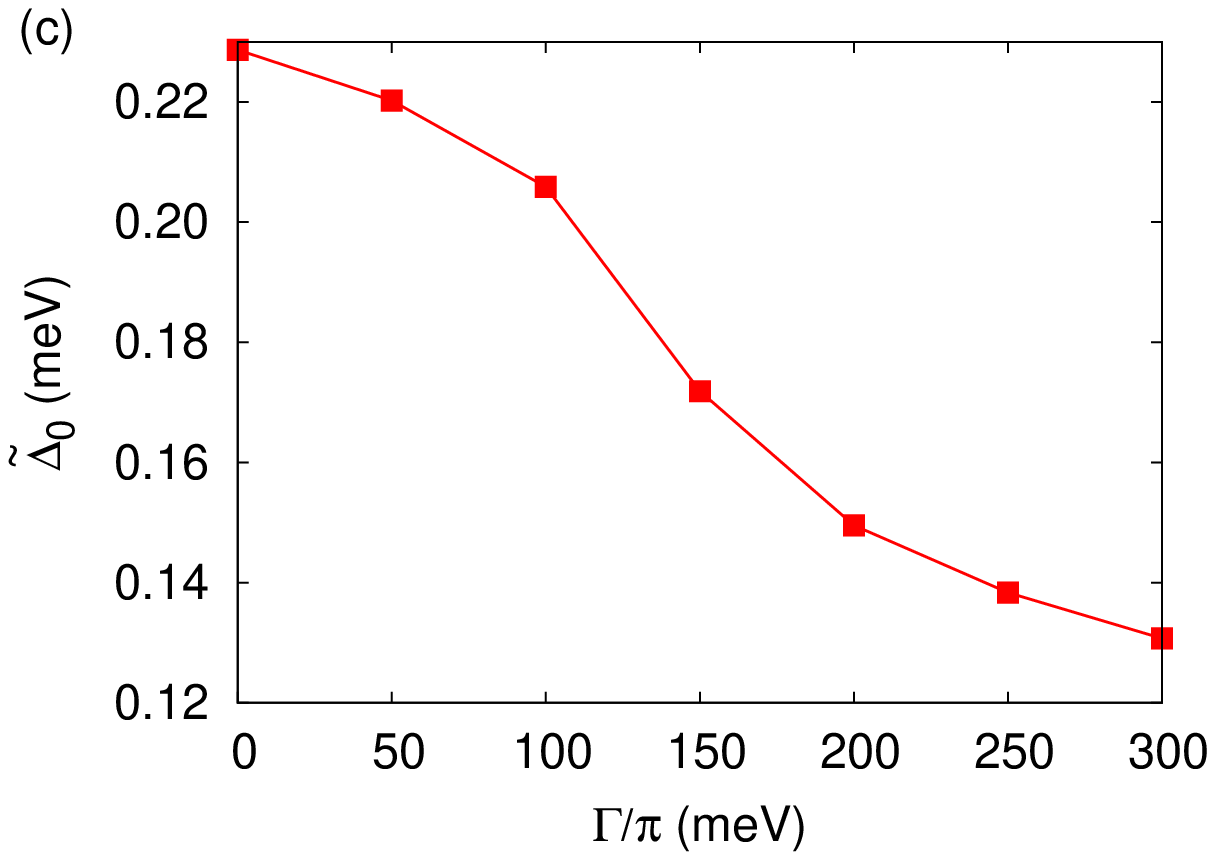} &
    \includegraphics[width=0.49\linewidth]{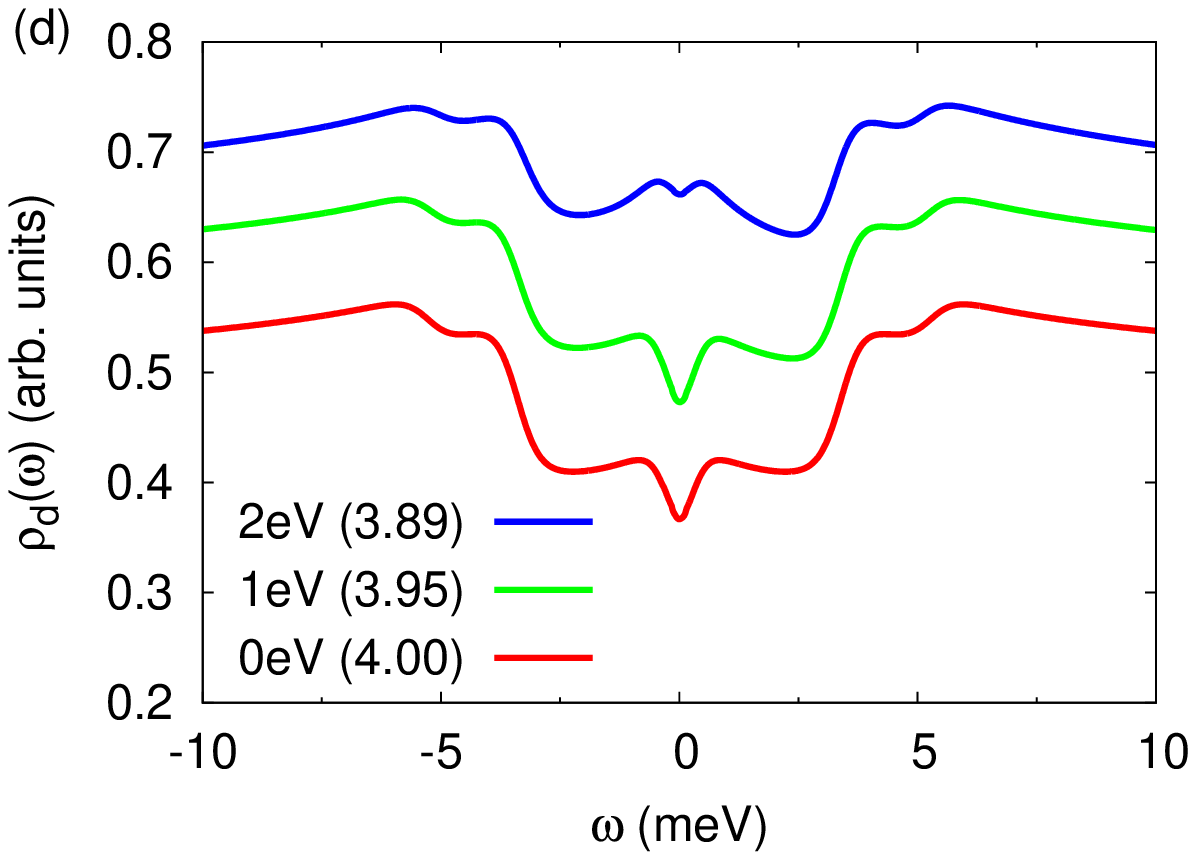} 
    \end{tabular}
  \caption{
    \label{fig:spin-2}
    Results for spin $S=2$ with $D=-1.55$meV and $E=0.35$meV, and temperature $kT=0.2$meV.
    (a) Spectral function $\rho_d(\omega)$ for different values of single-particle broadening $\Gamma$.
    (b) Temperature dependence of spectral function for $\Gamma/\pi=100$meV.
    (c) QST $\tilde\Delta_0$ as a function of single-particle broadening $\Gamma$.
    (d) Effect of charge fluctuations: spectra for $\Gamma/\pi=100$meV for 
    different energy level shifts $\delta\epsilon_d$. 
    In parenthesis the total occupancy $N_d$ of the impurity levels is given.  
  }
\end{figure}

The competition between Kondo effect and QST is also present for higher integer spin systems.
In Fig.~\ref{fig:spin-2} we show results for spin $S=2$, and magnetic anisotropy parameters 
$D=-1.55$meV and $E=0.35$meV chosen to reproduce spectra for Fe on Cu$_2$N \cite{Hirjibehedin07}.
The energy level diagram, and the spin composition of the corresponding states, are shown in 
Fig.~\ref{fig:method}. Here the QST splits both the $m=\pm2$ ground state doublet, by an amount 
$\Delta_0 \propto \frac{E^2}{D}$ as well as the $m=\pm 1$ doublet by $2E$.  Thus, 
the magnetic anisotropy completely lifts the degeneracy of the GS quintuplet leading to
step features in the calculated spectral functions [Fig.~\ref{fig:spin-2}(a)] that, for small 
values of $\Gamma$, resemble those measured for Fe on Cu$_2$N \cite{Hirjibehedin07},
and correspond to inelastic spin excitations from the ground state.
As $\Gamma$ is increased, the inelastic spin excitation step features both move 
to lower energies and become broader, indicating both the renormalization of the associated 
excitation energies and the decrease of the lifetimes due to the exchange coupling.  
As in the $S=1$ case, the step features associated with QST develop a triangular shape 
with increasing $\Gamma$ and finally turn into a split-Kondo peak.  
As $\Gamma$ is increased, the effective QST $\tilde{\Delta}_0$ is renormalized but remains finite
in the range of physically reasonable values of $\Gamma$ (up to 1eV) considered here [see Fig.~\ref{fig:spin-2}(c)].
We note that generally the Kondo coupling is also weaker for $S=2$ than for $S=1$ since
higher order processes are necessary to screen the spin.
For a somewhat smaller $E$ (not shown) the splitting of the Kondo peak can also 
vanish for spin $S=2$. 
Fig.~\ref{fig:spin-2}(b) shows the temperature dependence of the spectrum
for the case of relatively small coupling $\Gamma$. As the temperature is lowered, initially flat 
steps appear in the spectrum, which become increasingly triangular as the temperature is lowered
further.
As in the case of spin $S=1$, increasing the valence fluctuations by detuning 
the system from ph symmetry has a similar effect as increasing $\Gamma$ 
[see Fig.~\ref{fig:spin-2}(d)],
leading to a red-shift of the spin excitation energies. 
The detuning from ph symmetry also leads to an up-bending of the initially flat 
step features associated with QST and to the conversion into a split-Kondo peak.

The picture that emerges from our calculations is the following.  An atom with integer spin $S$, described with Hamiltonian (\ref{eq:imp}),  is  a closed quantum system   whose  quantum ground state has a built-in coherence between the 
 two classical ground states, with $S_z=\pm S$,  associated to the QST splitting  $\Delta_0$, that  would determine the frequency of the Rabi flops of the magnetization if the atom was initially prepared in an eigenstate of $S_z$.  As the coupling to the surface electrons is turned on, the atomic spin behaves like a quantum open system.   This results both in the  renormalization (reduction) of this Rabi frequency as well as spin relaxation, leading to a broadening of the steps. For sufficiently large coupling the QST splitting can be completely quenched, in line with previous results \cite{Delgado-2015}, and quantum coherence  between the states with opposite $S_z$ is lost. At that point Kondo screening takes over, and the zero bias Kondo feature appears.		
 
Interestingly, the evolution between the strong coupling Kondo regime and the weak coupling with step-wise excitations is continuous. 
At intermediate couplings the spectral functions show a split Kondo peak, that resembles the Zeeman split Kondo peak, but is driven 
by QST instead. The strength of the Kondo coupling is controlled both by $\Gamma$ and by the departure from the 
e-h symmetry point, that we change by tuning $\delta \epsilon_d$. Importantly, both the weak coupling picture with step-wise excitations 
associated with spin transitions, and the strong coupling Kondo regime, are also obtained in the case when the charge on the atom is not 
quantized, as suggested by DFT calculations \cite{Ferron2015,Panda2015}.

\section{Conclusions}

In summary, we have studied the competition between two important physical phenomena that affect integer spins, namely QST and Kondo 
screening. Our calculations permit to trace the evolution from the weak Kondo coupling regime, 
where the  stepwise $dI/dV$ spectra are renormalized,  resulting in the shift and the broadening  of the spin excitation energies, to the strong coupling regime, where the zero bias Kondo peak appears. 
This  accounts for  several   experimental observations \cite{Oberg14,Jacobson15}.
For  strong  Kondo coupling, QST can   be completely  quenched. Importantly, for the intermediate coupling regime  we  predict  a new physical effect: for $B=0$ an energy split Kondo peak can arise, because of quantum spin tunneling splitting, in analogy with the $B\neq0$ Zeeman split Kondo peak, recently investigated more closely by spin-polarized STM.\cite{Bergmann}

\section*{Author contribution statement}
JFR proposed to study the effect of Kondo screening on quantum spin tunneling
in integer spin systems. 
DJ implemented the solution of the Anderson model including the anisotropy term 
within the OCA code and performed all numerical calculations. 
Both authors contributed equally to the discussion and physical interpretation of the
results and to the writing of the manuscript.

\begin{acknowledgments}
JFR acknowledges financial support by MEC-Spain (FIS2013 47328 C2 2 P)
and Generalitat Valenciana (ACOMP 2010 070), Prometeo. 
We thank L. Glazman, R. \v{Z}itko and C. Hirjibehedin for fruitful discussions
during SPICE workshop ``Magnetic ad\-atoms as building blocks for quantum magnetism''. 
\end{acknowledgments}

\appendix

\section{One-Crossing Approximation}
\label{App:OCA}

The starting point are the eigenstates $\ket{m}$ and corresponding eigenenergies $E_m$ 
of the \emph{isolated }impurity Hamiltonian (\ref{eq:Himp_diag}). 
In terms of the impurity eigenstates $\ket{m}$ we can rewrite the hybridization term ${\cal V}_{\rm hyb}$ as
\begin{equation}
  {\cal V}_{\rm hyb} = \sum_{m,n}\sum_{k,\alpha,\sigma} V_{k\alpha} 
  \left(\ket{m}\bra{m}d_{\alpha\sigma}^\dagger\ket{n}\bra{n} c_{k\alpha\sigma} + {\rm h.c.} \right)
  \label{eq:Vhyb_mb}
\end{equation}
${\cal V}_{\rm hyb}$ connects eigenstates of ${\cal H}_{\rm imp}$ with different occupation numbers, i.e. $N_m=N_n\pm1$.
It is the fluctuations between the GS and excited states of the atom give rise to the Kondo effect. Since the 
electron-electron interaction on the impurity ($U$ and $J_H$) is generally large compared to the hybridization ($V_{k\alpha}$),
a perturbative treatment in terms of the latter is justified. 

We are now going to develop a perturbation theory for the many-body eigenstates $\ket{m}$ of ${\cal H}_{\rm imp}$
in terms of the hybridization term ${\cal V}_{\rm hyb}$. 
In order to proceed we associate so-called pseudo-particle (PP) field operators $a_m^\dagger$, $a_m$ 
with the many-body eigenstates $\ket{m}$ of ${\cal H}_{\rm imp}$:
\begin{equation}
  \label{eq:pps}
  \ket{m} = a_m^\dagger \ket{\tilde{0}} \hspace{1ex}\mbox{and}\hspace{1ex} a_m \ket{m} = \ket{\tilde{0}}
\end{equation}
where $\ket{\tilde{0}}$ is the PP vacuum. The auxiliary fields $a_m^\dagger$ and $a_m$ 
obey either commutation or anticommutation rules depending on whether the corresponding many-body
state is bosonic (even number of electrons) or fermionic (odd number of electrons):
\begin{equation}
  \label{eq:pp_commutators}
  [a_m,a_m^\dagger] = 1 \hspace{1ex}\mbox{(Bosons)};\, \{a_m,a_m^\dagger\} = 1 \hspace{1ex}\mbox{(Fermions)}
\end{equation}
Note that $a_m^\dagger$ and $a_m$ are only auxiliary fields that {\it per se} do not have a 
physical meaning. In order to give this construction a physical significance one has to impose an 
additional constraint that enforces the conservation of the PP number to one, i.e. the system
can only be in one and only one state $\ket{m}$ at a time:
\begin{equation}
  \label{eq:pp_charge}
  Q = \sum_m a_m^\dagger a_m \equiv 1
\end{equation}

The PPs are related to the real electrons by:
\begin{equation}
  \label{eq:rel_d_pp}
  d_{\alpha\sigma} = \sum_{m,n} \bra{m} d_{\alpha\sigma} \ket{n} a_m^\dagger a_n
\end{equation}
In the PP picture the Hamiltonian of the isolated atom becomes
\begin{equation}
  \label{eq:Himp_pp}
  {\cal H}_{\rm imp} = \sum_m E_m a_m^\dagger a_m + \lambda\,\left(\sum_m a_m^\dagger a_m-1\right)
\end{equation}
where the last term is a Lagrangian constraint imposing the afore mentioned conservation 
of the PP charge $Q$. The Lagrange multiplier $\lambda$ can 
be seen as a (negative) chemical potential for the PPs. Imposing the constraint via
the Lagrange multiplier is done e.g. in the Slave-Boson Mean-Field Approximation of the 
Anderson model.\cite{SBMF} However, for the diagrammatic expansion it is more convenient
to work in the grand-canonical ensemble with respect to the PP charge $Q$.
The expectation values of a physical observable $A$ can be calculated in the grand canonical
PP ensemble and then projected to the physical subspace ($Q=1$) via the Abrikosov trick:\cite{Abrikosov1965}
\begin{equation}
  \label{eq:Abrikosov}
  \langle A \rangle = \lim_{\lambda\rightarrow\infty} \frac{ \langle Q A \rangle_\lambda}{\langle Q \rangle_\lambda}
\end{equation}

In the PP picture the hybridization term (\ref{eq:Vhyb_mb}) becomes:
\begin{equation}
  \label{eq:Vhyb_pp}
  {\cal V}_{\rm hyb} = \sum_{m,n}\sum_{k,\alpha,\sigma} V_{k\alpha} 
  \left( D^{\alpha\sigma\dagger}_{mn} a_m^\dagger a_n c_{k\alpha\sigma} + D^{\alpha\sigma}_{mn} c_{k\alpha\sigma}^\dagger a_m^\dagger a_n \right)
\end{equation}
where we have introduced the matrix elements $D_{mn}^{\alpha\sigma}=\bra{m}d_{\alpha\sigma}\ket{n}$ and 
$D_{mn}^{\alpha\sigma\dagger}=\bra{m}d^\dagger_{\alpha\sigma}\ket{n}$.
The hybridization term ${\cal V}_{\rm hyb}$ is now treated as a perturbation to the 
Hamiltonian of the uncoupled impurity and bath ${\cal H}_0={\cal H}_{\rm imp} + {\mathcal H}_{\rm bath}$. 
The introduction of PPs obeying (anti-) commutation relations allows us to make
use of the machinery of quantum field theory to develop a diagrammatic perturbation expansion
since Wick's theorem applies.

We now introduce the bare PP propagators $G_m^0(\tau)=-\langle T_\tau a_m(\tau) a_m^\dagger(0)\rangle_0$ 
which will be denoted by dashed lines:
\begin{equation}
  \label{eq:pp_gf0}
  G_m^0(i\omega) = \frac{1}{i\omega-\lambda-E_m} = \includegraphics[width=0.2\linewidth,raise=-2.5ex]{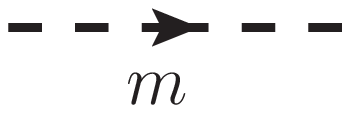}
\end{equation}
The bath electron propagator is denoted by a full line:
\begin{equation}
  \label{eq:bath_gf}
  g_{k\alpha\sigma}(i\omega) = \frac{1}{i\omega-\epsilon_{k\alpha}}
  = \includegraphics[width=0.2\linewidth,raise=-2.5ex]{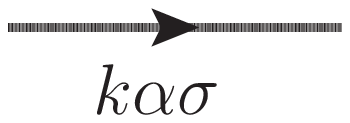}
\end{equation}

Expanding the full PP propagator $G_m(\tau)=\langle T_\tau a_m(\tau) a_m^\dagger(0)\rangle$ in terms of the 
hybridization ${\cal V}_{\rm hyb}$, we can integrate out the bath electrons, and end up with an 
effective retarded two-particle interaction between PPs mediated by the bath electrons:
\begin{eqnarray}
  \label{eq:pp_interaction}
  \lefteqn{\tilde{\cal V}_{\rm hyb}(\tau-\tau^\prime) = \sum_{\alpha,\sigma,m,n,m^\prime,n^\prime} 
  D^{\alpha\sigma\dagger}_{mn} \, D^{\alpha\sigma}_{n^\prime m^\prime} \; \times} \\ 
  && \times \; a_m^\dagger(\tau) a_n(\tau) 
  \left[ \sum_k |V_{k\alpha}|^2 g_{k\alpha\sigma}(\tau-\tau^\prime) \right] 
  a_{n^\prime}^\dagger(\tau^\prime) a_{m^\prime}(\tau^\prime)
  \nonumber
\end{eqnarray}
The term in square brackets is called the hybridization function $\Delta_{\alpha}(\tau-\tau^\prime)$
which in the real frequency domain is given by eq.~(\ref{eq:hybfunc}).

The full PP propagators are denoted by double dashed lines.
In terms of a PP self-energy $\Sigma_m(\omega)$ 
which captures the interaction with other PPs via ${\cal V}_{\rm hyb}$,
the full propagator can be written as
\begin{equation}
  \label{eq:full_pp_gf}
  G_m(i\omega) = \frac{1}{i\omega-\lambda-E_m-\Sigma_m(i\omega)}
  = \includegraphics[width=0.2\linewidth,raise=-2.5ex]{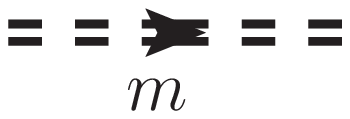} 
\end{equation}

In lowest order (2nd order in $V_{k\alpha}$ or 1st order $\Delta_\alpha$) 
the PP self-energy is given by the following diagrams:
\begin{equation}
  \label{eq:pp_self1}
  \Sigma_m(\omega) = 
  \includegraphics[width=0.2\linewidth,raise=-2.5ex]{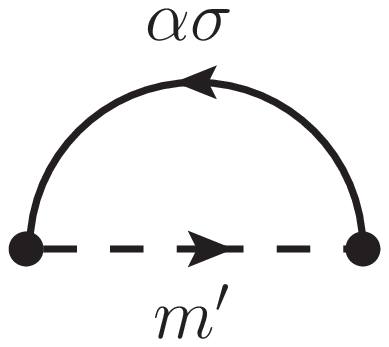} + \includegraphics[width=0.2\linewidth,raise=-2.5ex]{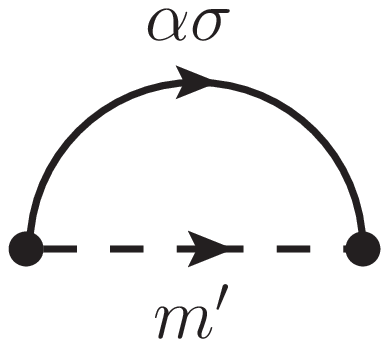}
\end{equation}
The conduction electron line (full lines) $\alpha\sigma$ correspond
to the hybridization function  $\Delta_\alpha(\omega)$.

The first diagram (``backward'' diagram) corresponds to adding an electron to the impurity site, i.e. $N_{m^\prime}=N_m+1$
and the second diagram (``forward'' diagram) to removing an electron, i.e. $N_{m^{\prime\prime}}=N_m-1$.
The non-crossing approximation (NCA) consists in an infinite resummation of these lowest order diagrams
where conduction electron lines do not cross (hence the name). Replacing the bare propagators by full 
propagators in the above self-energy diagrams, one obtains the NCA self-energy diagrams:
\begin{eqnarray}
  \label{eq:pp_self_nca}
  \lefteqn{
    \Sigma_m^{\rm NCA}(\omega) = 
    \includegraphics[width=0.2\linewidth,raise=-2.5ex]{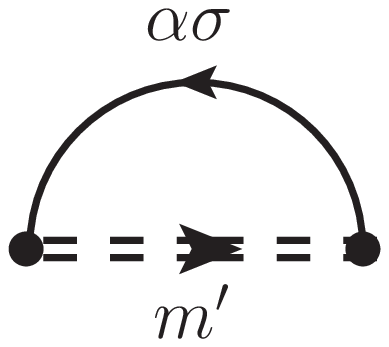} + \includegraphics[width=0.2\linewidth,raise=-2.5ex]{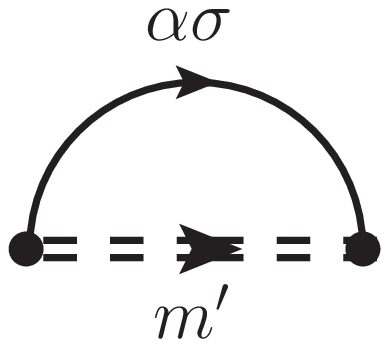}
  }
  \nonumber\\
  &=& -\sum_{m^\prime\alpha\sigma} \int\frac{d\nu}{\pi} 
  \left[ 
    |D_{mm^\prime}^{\alpha\sigma}|^2 \; f(\nu) \; \Gamma_\alpha(\nu) \; G_{m^\prime}(\omega+\nu) \right. \nonumber\\
    && \hspace{0.05\linewidth} + \left. |D_{m^\prime{m}}^{\alpha\sigma}|^2 \; f(-\nu) \; \Gamma_\alpha(\nu) \; G_{m^\prime}(\omega-\nu)
    \right]
\end{eqnarray}
where $\Gamma_\alpha(\omega)\equiv-\Im\,\Delta_\alpha(\omega)$ is the single-particle broadening of the impurity levels
$\alpha$ by the coupling to the conduction electron bath, and $f(\nu)$ is the Fermi function. Hence the NCA self-energy
for a PP $m$ is given by a convolution of the imaginary part of the hybridization function $\Gamma_\alpha$ 
with the propagators of all other PPs $m^\p$ that $m$ is interacting with via the conduction 
electron bath.

The OCA diagrams are second order in $\Delta_\alpha(\omega)$ and involve crossing conduction electron lines:
\begin{eqnarray}
  \label{eq:pp_self_oca}
  \lefteqn{\Sigma_m^{\rm OCA}(\omega) = \Sigma_m^{\rm NCA}(\omega) }\nonumber\\
  && + \includegraphics[width=0.4\linewidth,raise=-2ex]{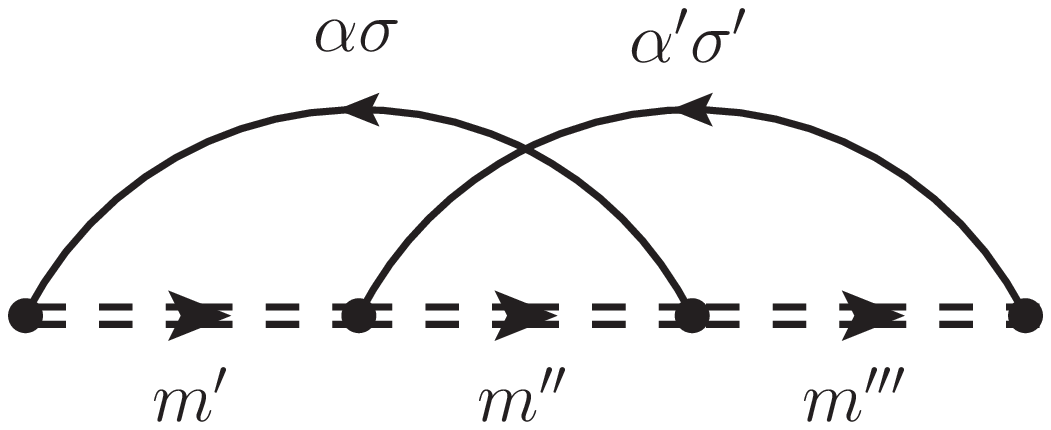} 
     + \includegraphics[width=0.4\linewidth,raise=-2ex]{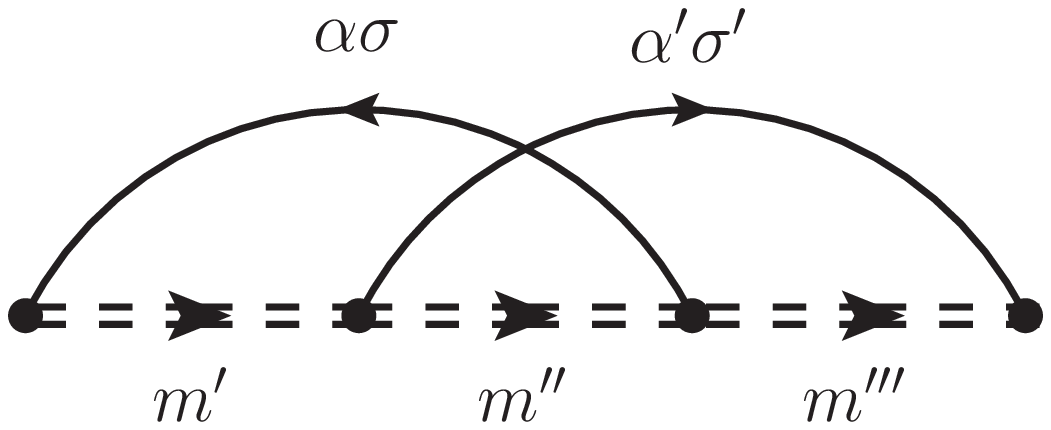}\nonumber\\
  && + \includegraphics[width=0.4\linewidth,raise=-2ex]{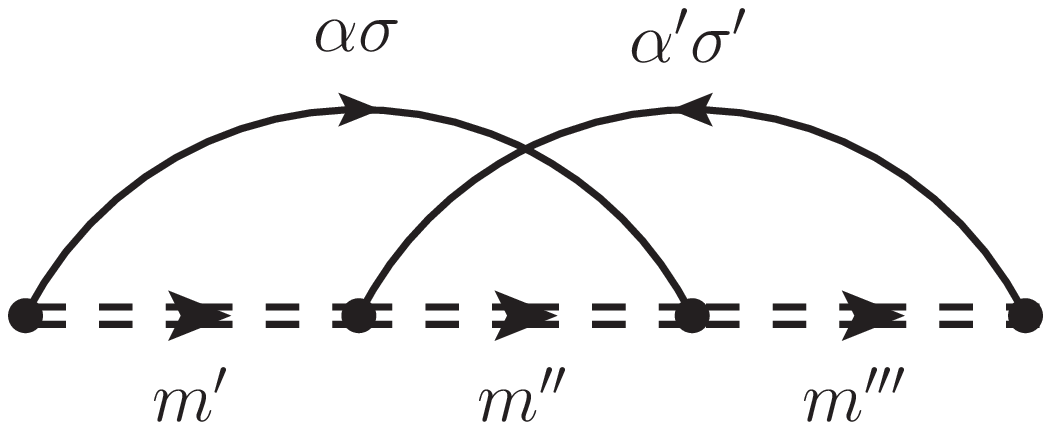} 
     + \includegraphics[width=0.4\linewidth,raise=-2ex]{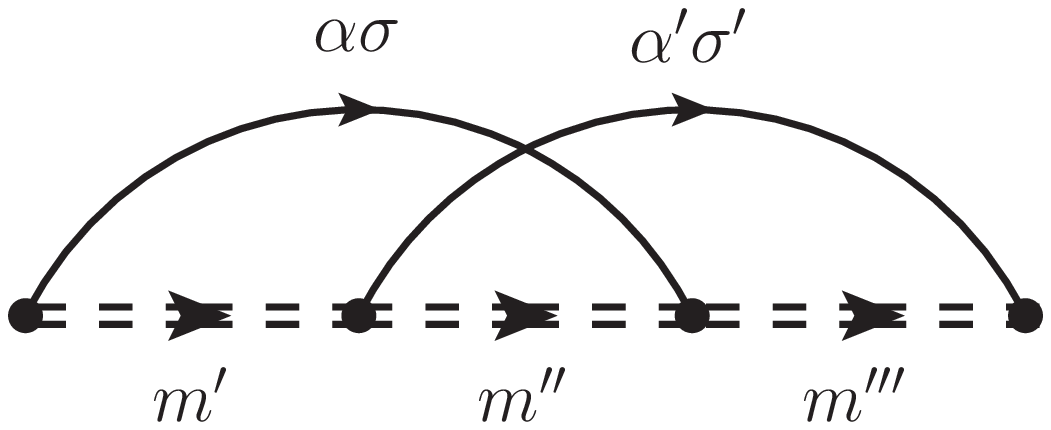}\nonumber \\
\end{eqnarray}
The algebraic expressions for the OCA self-energy are much more complicated than the NCA ones,
and involve double convolutions of two hybridization functions with three PP propagators.
The exact expressions can be found in the literature.\cite{Haule:2001,Haule:2010}

The NCA/OCA  equations are a set of coupled integral equations that 
have to be solved {\it self-consistently} since the self-energy of a PP $m$ depends
on the {\it full} propagators of other PPs $m^\prime$.
Once the NCA/OCA equations are solved, the real impurity electron propagator can be determined
by making use of the relation (\ref{eq:rel_d_pp}). Hence the real electron propagator can be
calculated from a two-particle correlation function for PPs:
\begin{eqnarray}
  \label{eq:real_gf}
  G_{\alpha\sigma}(\tau) &=& -\langle T_\tau d_{\alpha\sigma}(\tau) d_{\alpha\sigma}^\dagger(0) \rangle
  \nonumber\\
  &=& -\sum_{m,n,m^\prime,n^\prime} D_{mn}^{\alpha\sigma\dagger} D_{nm}^{\alpha\sigma}
  \langle T_\tau a_m^\dagger(\tau) a_n(\tau) a_{n}^\dagger a_{m} \rangle
  \nonumber\\
\end{eqnarray}
In NCA the vertex correction $\Lambda$ is neglected.
It is then found that the real electron spectral function $\rho_{\alpha\sigma}(\omega)=-\Im{G_{\alpha\sigma}(\omega)}/\pi$
can be calculated from a convolution of PP spectral functions $A_m(\omega)=-\Im\,G_m(\omega)/\pi$:
\begin{eqnarray}
  \label{eq:rho_d}
    \rho_\alpha(\omega) &=& \frac{1}{\langle{Q}\rangle_\lambda} \sum_{mm^\prime} \int d\varepsilon\,
    e^{-\beta\varepsilon}\;[1+e^{-\beta\omega}] \times
  \nonumber\\
  && \hspace{0.05\linewidth} \times |D^{\alpha\sigma}_{mm^\prime}|^2 \; A_m(\varepsilon) \; A_{m^\prime}(\omega+\varepsilon)
\end{eqnarray}
where $Q$ is the PP charge which can be calculated directly from the PP 
spectral functions:
\begin{equation}
  \langle Q \rangle = \int d\omega \, e^{-\beta\omega} \sum_m A_m(\omega)  
\end{equation}
Again the corresponding expression for calculating the real electron spectral function 
within OCA is much more complicated as it involves double convolutions of PP
correlation functions and the hybridization function.\cite{Haule:2001,Haule:2010}


\begin{thebibliography}{0}

\bibitem{Gatteschi-book} D. Gatteschi, R. Sessoli, and J. Villain, Molecular Nanomagnets (Oxford Univ. Press, Oxford, 2006) 

\bibitem{Abragam-Bleaney}  A. Abragam and B. Bleaney, {\em Electron Paramagnetic Resonance of Transition Ions} (Oxford Univ. Press, Oxford, 2012)

\bibitem{Hewson-book}  A.~C. Hewson, {\em The Kondo Problem to Heavy Fermions} (Cambridge Univ. Press, Cambridge, 1997)

\bibitem{Hirjibehedin07} C. Hirjibehedin, C-Y Lin. A.F. Otte, M. Ternes, C. P. Lutz, B. A. Jones, and A. J. Heinrich, 
  Science {\bf 317}, 1199 (2007)  

\bibitem{Khajetoorians2010} A. A. Khajetoorians, B. Chilian, J. Wiebe, S. Schuwalow, F. Lechermann, and R.  Wiesendanger, 
  Nature {\bf 467}, 1084 (2010)
  
\bibitem{FePc} N. Tsukahara, K. I. Noto, M. Ohara, S. Shiraki, N. Takagi, Y. Takata, J. Miyawaki, M. Taguchi, A. Chainani, 
  S. Shin, and M. Kawai, Phys. Rev. Lett. {\bf 102},  167203 (2009)

\bibitem{Sasaki} S. Sasaki, S. De Franceschi, J. M. Elzerman, W. G. van der Wiel, M. Eto, S. Tarucha,  and L. P. Kouwenhoven, 
  Nature {\bf 405}, 764 (2000) 

\bibitem{Parks2010} J. J. Parks,  A. R. Champagne, T. A. Costi, W. W. Shum, A. N. Pasupathy, 
  E. Neuscamman, S. Flores-Torres, P. S. Cornaglia, A. A. Aligia, C. A. Balseiro, G. K.-L. Chan, 
  H. D. Abru\~na, and D. C. Ralph, Science {\bf 328}, 1370 (2010)

\bibitem{Tsukahara2011}
  N. Tsukahara, S. Shiraki, S. Itou, N. Ohta, N. Takagi, and M. Kawai,  Phys. Rev. Lett. {\bf 106}, 187201 (2011)

\bibitem{Mugarza12}
  A. Mugarza, R. Robles, C. Krull, R. Korytar, N. Lorente, and P. Gambardella
  Phys. Rev. B {\bf 85}, 155437  (2012)
  
\bibitem{Ho} Y. Jiang, Y. N. Zhang, J. X. Cao, R. Q. Wu, and W. Ho, Science {\bf 333}, 324 (2011) 

\bibitem{Madhavan98} V. Madhavan, W. Chen, T. Jamneala, M. F. Crommie, and N. S. Wingreen, Science {\bf 280}, 567 (1998)


\bibitem{Garg93} A. Garg,  Europhys. Lett. {\bf 22},  205 (1993)

\bibitem{Sessoli-Wernsdorfer}  R. Sessoli and W. Wernsdorfer,  Science {\bf 284}, 133 (1999)

\bibitem{Delgado12} F. Delgado,  J. Fern\'andez-Rossier, Phys. Rev. Lett. {\bf 108 } 196602 (2012)

\bibitem{Delgado-2015} F. Delgado,  S. Loth, M. Zielinski, and J. Fern\'andez-Rossier, EPL {\bf 109}, 57001 (2015)

\bibitem{Oberg14} J. Oberg {\em et al.}, Nature Nanotech. {\bf 9}, 64 (2014)

\bibitem{Delgado-Surf-Sci} F. Delgado, C. F. Hirjibehedin, and J. Fern\'andez-Rossier, Surf. Sci. {\bf 630}, 337 (2014) 

\bibitem{Otte08} A. F. Otte, M. Ternes, K. von Bergmann, S. Loth, H. Brune, C. P. Lutz, 
  C. F. Hirjibehedin, and A. J. Heinrich, Nature Physics {\bf 4}, 847 (2008) 

\bibitem{Jacobson15} P. Jacobson, T. Herden, M. Muenks, G. Laskin, O. Brovko, V. Stepanyuk, M. Ternes, and K. Kern, arXiv:1505.02277.

\bibitem{JFR09} J. Fern\'andez-Rossier, Phys. Rev. Lett.  {\bf 102}, 256802  (2009)

\bibitem{Zitko} R. \v{Z}itko and Th. Pruschke, New. J. Phys. {\bf 12}, 063040 (2010); 
  R. \v{Z}itko, R. Peters, and Th. Pruschke, Phys. Rev. B {\bf 78}, 224404 (2008)

\bibitem{Sanvito} A. Hurley, N.   Baadji, and S.  Sanvito, Phys. Rev. B {\bf 84}, 115435 (2011) 

\bibitem{Ternes} M. Ternes, New Journal of Physics {\bf 17},  063016  (2015)

\bibitem{Ferron2015}  A. Ferr\'on,  J. L. Lado,  and J. Fern\'andez-Rossier, Phys. Rev. B{\bf 92} 174407 (2015)

\bibitem{Panda2015} S. K. Panda, I. Di Marco, O. Gr\r{a}n\"as, O. Eriksson, and J. Fransson, arXiv:1511.07909

\bibitem{Datta-book} S. Datta, \emph{Electronic Transport in Mesoscopic Systems} (Cambridge Univ. Press, Cambridge, 1995)

\bibitem{Ujsaghy2000} O. \'Ujs\'aghy, J. Kroha, L. Szunyogh, and A. Zawadowski, Phys. Rev. Lett. {\bf 85}, 2557 (2000)

\bibitem{Zitko11} R. \v{Z}itko, O. Bodensiek, and Th. Pruschke, Phys. Rev. B {\bf 83}, 054512 (2011)

\bibitem{Korytar12} R. Koryt\'ar, N. Lorente, and J.-P. Gauyacq, Phys. Rev. B {\bf 85}, 125434 (2012)

\bibitem{Xue08} X. Chen, Y.-S. Fu, S.-H. Ji, T. Zhang, P. Cheng, X.-C. Ma, X.-L. Zou, W.-H. Duan, J.-F. Jia, and Q.-K. Xue, 
  Phys. Rev. Lett. {\bf 101}, 197208 (2008)

\bibitem{adatom_hyb} B. Surer, M. Troyer, Ph. Werner, T.~O. Wehling, A. M. L\"auchli, A. Wilhelm, and A.~I. Lichtenstein,
  Phys. Rev. B {\bf 85}, 085114 (2012); D. Jacob, J. Phys.: Condes. Matter {\bf 27}, 245606 (2015)

\bibitem{Pruschke} T. Pruschke and N. Grewe, Z. Phys. B {\bf 74}, 439 (1989);
\bibitem{Haule:2001} K. Haule, S. Kirchner, J. Kroha, and P. W\"olfle, Phys. Rev. B {\bf 64}, 155111 (2001)
\bibitem{Haule:2010} K. Haule, C. H. Yee, and K. Kim, Phys. Rev. B {\bf 81}, 195107 (2010)
\bibitem{OCA-artifacts} 
  T. A. Costi, and J. Kroha, and P. W\"olfle, Phys. Rev. B {\bf 53}, 1850 (1996); 
  N. Grewe, S. Schmitt, T. Jabben and F. B. Anders, J. Phys.: Condens. Matter {\bf 20}, 365217 (2008)

\bibitem{NRG} R. Bulla, T. A. Costi, and T. Pruschke, Rev. Mod. Phys. {\bf 80}, 395 (2008)

\bibitem{Ruegg13} A. R\"uegg, E. Gull, G.~A. Fiete, and A.~J. Millis, Phys. Rev. B {\bf 87}, 075124 (2013)

\bibitem{Jacob13} D. Jacob, M. Soriano, and J.~.J. Palacios, Phys. Rev. b {\bf 88}, 134417 (2013)
\bibitem{Karan15} S. Karan, D. Jacob, M. Karolak, C. Hamann, Y. Wang, A. Weismann, A. I. Lichtenstein, R. Berndt, 
  Phys. Rev. Lett. {\bf 115}, 016802 (2015)

\bibitem{Romeike:2006} C. Romeike, M. R. Wegewijs, W. Hofstetter, and H. Schoeller, Phys. Rev. Lett. {\bf 97}, 206601 (2006) 

\bibitem{Zyazin:2010} A. S. Zyazin, J. W. G. van den Berg, E. A. Osorio, H. S. J. van der Zant, N. P. Konstantinidis,
  M. Leijnse, M. R. Wegewijs, F. May, W. Hofstetter, C. Danieli, and A. Cornia, Nano Lett. {\bf 10}, 3307 (2010)

\bibitem{valence-fluctuations} T. A. Costi, A. C. Hewson, and V. Zlati\'c, J. Phys.: Condens. Matter {\bf 6}, 2519 (1994)

\bibitem{Bergmann} K. von Bergmann, M. Ternes, S. Loth, C. P. Lutz, and A. J. Heinrich, Phys. Rev. Lett. {\bf 114}, 076601 (2015)

\bibitem{SBMF} D. M. Newns and N. Read, J. Phys. C {\bf 16}, 3273 (1983)
\bibitem{Abrikosov1965} A. A. Abrikosov, Physica {\bf 2}, 21 (1965)
\bibitem{Coleman:1984} P. Coleman, Phys. Rev. B {\bf 29}, 3035 (1984)
\bibitem{Kroha:1998} J. Kroha and P. W\"olfle, Acta Phys. Pol. B {\bf 29}, 3781 (1998) 

\bibitem{Nevidomskyy2009} A. H. Nevidomskyy and P. Coleman, Phys. Rev. Lett. {\bf 103}, 147205 (2009)

\end{thebibliography}
\end{document}